\documentclass[10pt,twocolumn,letterpaper]{article}

\usepackage[pagenumbers]{cvpr}

\usepackage{colortbl}
\usepackage{float}
\definecolor{bifocallightblue}{HTML}{DDF3FF}

\definecolor{preprintlink}{rgb}{0.21,0.49,0.74}
\usepackage[pagebackref,breaklinks,colorlinks,allcolors=preprintlink]{hyperref}

\title{BiFocal: Self-Evolving Local-Global Memory Agents for Long-Video Reasoning}

\author{Jiangwen Dong\\
The Hong Kong Polytechnic University\\
{\tt\small jiangwen.dong@connect.polyu.hk}
\and
Hao Lan\\
Tsinghua University\\
{\tt\small lanhao@tsinghua.edu.cn}
\and
Wanyu Lin\\
The Hong Kong Polytechnic University\\
{\tt\small wan-yu.lin@polyu.edu.hk}
}

\begin{document}
\maketitle
\begin{abstract}
Long-video reasoning is challenging because a video is long and redundant, while the evidence for a question is sparse and spread over time. Memory agents are a promising multi-agent answer: a team of agents, each reading one part of the video, can keep useful observations and reuse them for later questions. Existing memory designs, however, focus on search or storage and keep one shared store for the whole video. Such a store does not record which agent wrote an entry or whether the entry was checked against the video, so one policy must learn both what to keep and what to share, and unreliable entries are reused. We argue that memory should be organized by role and scope, and that agents should learn to build, check, and evolve it. We propose \textbf{BiFocal}, a multi-agent system with local and global memory. Each Reader agent watches one part of the video and keeps detailed notes as local memory. These notes are ordered by time into a shared global memory. The orchestrator, which decides what to read and answers, checks a global entry against its source video frames before citing it. The Readers and the orchestrator learn separate policies from task outcome with role-specific costs. We train on the MLVU development split and test on LVBench, Video-MME, LongVideoBench, and the MLVU test split. At three matched call and frame budgets, BiFocal is more accurate than a fixed prompt policy with the same memory in all twelve benchmark-budget settings, by 1.2 to 9.1 points, with 14 to 20 percent fewer calls.
\end{abstract}

\begin{figure*}[t]
\centering
\includegraphics[width=\textwidth]{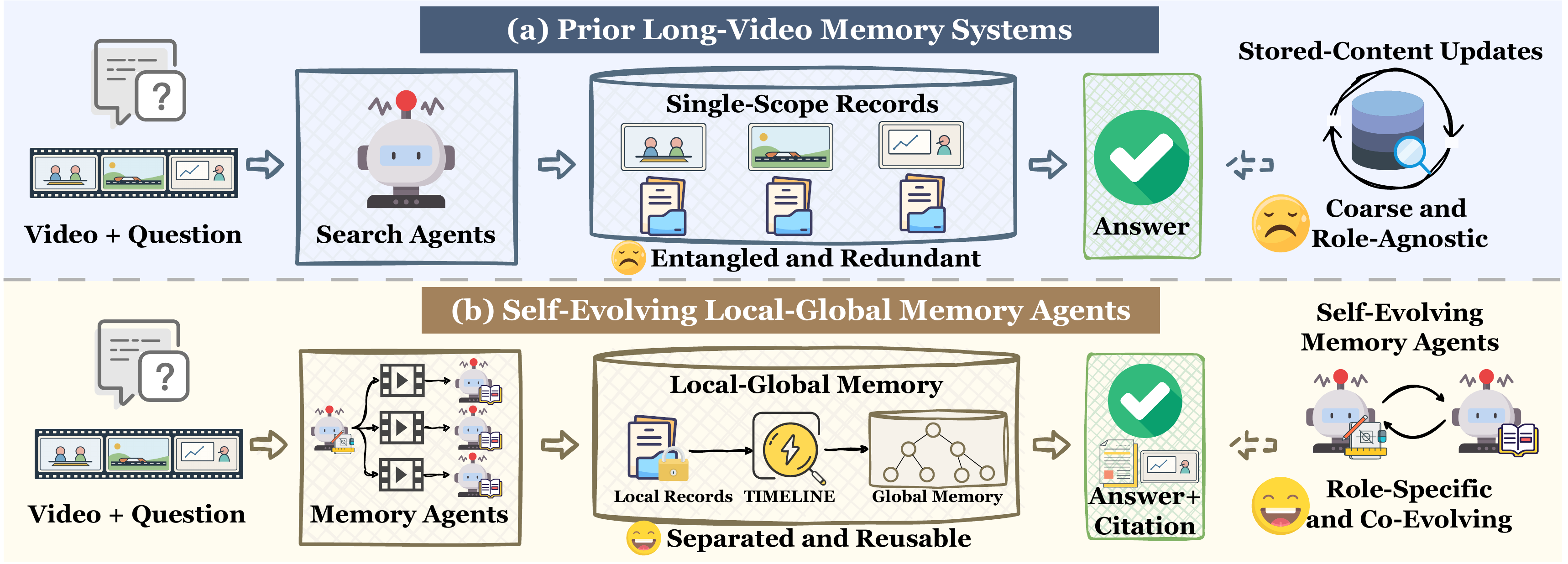}
\caption{\textbf{Memory and agent evolution in long-video reasoning (illustrative).} \emph{(a)} A typical prior design keeps one video-level memory with role-agnostic updates. \emph{(b)} BiFocal keeps local Reader records and a global team timeline, with separate policies for O and the Readers.}
\label{fig:comparison}
\end{figure*}

\section{Introduction}

Long-video reasoning is challenging because the evidence needed to answer a question is sparse. A video may run for hours, while a question may depend on only a few seconds. Due to context window or budget constraints, a model can make only a limited number of calls and inspect only a limited number of sampled frames, which cover a small fraction of the video~\cite{liu2026bandwidth}. A larger frame budget improves coverage, but it does not show which evidence matters. The system must decide where to look next, when to request more evidence, and when to stop. Across questions, an early fact may help several later questions, and videos often carry many (about 15 per video on LVBench and 3 on Video-MME), but re-reading every earlier segment wastes the budget, and keeping every observation adds noise. Reusing an observation is only safe if it can be traced back to its source frames. The system therefore needs a way to keep useful details without keeping everything.

Prior work attacks this budget problem by compressing hour-long videos into shorter representations~\cite{ren2025vamba,kim2026mambamia}, by searching for evidence while answering~\cite{wang2024videotree,zhang2025deepvideodiscovery,ding2026videozoomer,zhang2026vts}, or by assigning a team of agents to different video parts or responsibilities~\cite{chen2025lvagent,chen2026videochatm1,yan2026symphony,wang2026videohv,qiu2026videoseal,liu2026videomind,li2026lenswalk}. Coordination is not free~\cite{cemri2025mast}, and these systems usually coordinate within one question rather than keeping team memory for later questions. Memory-based methods instead reuse information later by storing events, objects, compact representations, or retrievable observations~\cite{fan2024videoagent,long2026m3agent,yeo2026worldmm,zuo2025videolucy,agarwal2026memstream,choi2026merit}, and recent work updates memory as questions arrive~\cite{wang2026memevo,lyu2026pivot}.
This setting differs from searching for one answer in one video. The team answers a sequence of questions about each video, every one under a fixed call and frame budget with memory persisting across them, and is judged on accuracy at equal caps, so the question is how the team allocates a limited budget across its agents and over a video's questions, and not only how many agents it uses.

However, multi-agent and memory research leave two gaps. Multi-agent systems usually coordinate only within a question, while memory systems usually let one agent manage a shared store. Neither line separates observations owned by one agent from facts reviewed for the whole team, so one policy must learn both what local details to keep and what facts to share. A local observation can be detailed but unreliable, whereas shared memory must keep enough source information for later checking. A Reader sees only its own part of the video and can keep rich notes, while the team needs a compact record that fits in one agent's context and still points back to the frames it came from. We argue that each agent should keep a local memory and the team a shared global memory. Detailed evidence should remain with the agent that observed it, while shared entries should retain their source time, version, and stored-frame handle. The roles should also have separate policies trained on the same task outcome with role-specific costs. Figure~\ref{fig:comparison} contrasts local and global memory with one shared store.

To address the problem, we propose \textbf{BiFocal}, a self-evolving local and global memory system for multi-agent long-video reasoning. A team of Reader agents inspects different parts of the video. Each Reader writes local records for its assigned video part, and each record keeps its source, version, and linked frames. An orchestrator (O) sees the question, the available memory, and the remaining budget. O requests reads and reviews local evidence against the stored frames. The harness builds a chronological global timeline from stored observations, and retrieval selects an entry and its source anchor before checking and answering. A non-learned harness enforces the budget and checks every memory operation, while O and the Readers improve through later training rounds. BiFocal thus evolves at two time scales: memory is updated within a video as questions arrive, and the policies are updated across training rounds, so the agents learn to build, check, and use memory over time.
Our contributions are:

\begin{itemize}
\item \textbf{Local and global memory for a reading team.} Each Reader keeps detailed local records of its own video part. The harness builds a chronological global timeline of the stored observations for the team, and the \textsc{review} action checks an entry against its stored frames before it is cited.
\item \textbf{Separate self-evolving policies.} O and the Readers are trained with separate adapters from task outcome minus a capped call and frame cost. O learns where to inspect and when to use or defer memory, and the Readers learn what to keep and report. A non-learned harness enforces the budget and validates every memory operation.
\item \textbf{Gains at matched budgets.} At equal call and frame caps, BiFocal is more accurate than the fixed prompt policy (same harness and memory, no training) on LVBench, Video-MME, MLVU test, and LongVideoBench in all twelve cells, by 1.2 to 9.1 points with 14 to 20\% fewer calls. The paired 95\% interval excludes zero in nine cells. Training uses only MLVU dev, so LVBench, Video-MME, and LongVideoBench test zero-shot transfer.
\end{itemize}

\section{Related Work}

\subsection{Agentic AI for Long-Video Reasoning}
\textit{Efficient video models.} VAMBA processes hour-long videos with a hybrid Mamba-Transformer~\cite{ren2025vamba}, and MambaMia compresses long video tokens with hierarchical state-space blocks~\cite{kim2026mambamia}. \textit{Search agents.} Deep Video Discovery searches a video with planner-guided tools~\cite{zhang2025deepvideodiscovery}, VideoZoomer learns temporal focusing with reinforcement learning~\cite{ding2026videozoomer}, and Video Tree Search explores a temporal tree and self-corrects its evidence path~\cite{zhang2026vts}. \textit{Multi-agent systems.} VideoChat-M1 trains multiple agents to plan collaboratively~\cite{chen2026videochatm1}, Symphony coordinates specialized agents~\cite{yan2026symphony}, and VideoHV-Agent pairs hypothesis generation with cross-agent verification~\cite{wang2026videohv}. LVAgent uses multi-round collaboration among multimodal agents~\cite{chen2025lvagent}, VideoSEAL separates a planner from an evidence-backed inspector~\cite{qiu2026videoseal}, VideoMind links role-specific LoRA agents for temporal reasoning~\cite{liu2026videomind}, and LensWalk separates a reasoner from an active observer~\cite{li2026lenswalk}. These methods improve evidence access, but their coordination is generally episode-local.

\subsection{Memory for Long-Video Agents}
\textit{Memory structure and retrieval.} VideoAgent stores temporal events and object states in a unified memory~\cite{fan2024videoagent}, and VideoLucy backtracks through coarse-to-fine memory~\cite{zuo2025videolucy}. M3-Agent builds entity-centric episodic and semantic memory~\cite{long2026m3agent}, and WorldMM combines episodic, semantic, and visual memories with adaptive retrieval~\cite{yeo2026worldmm}. MemStream maintains a dynamic KV-cache memory for video streams~\cite{agarwal2026memstream}, and MERIT retrieves multi-key episodic memories for ultra-long video~\cite{choi2026merit}. \textit{Evolving and multi-question memory.} MEMEVO evolves memory across queries~\cite{wang2026memevo}, MAGIC-Video builds a structured memory graph with narrative chains~\cite{li2026magicvideo}, and VideoSleuth organizes narrative elements through on-demand tools~\cite{li2026videosleuth}. PIVOT handles multiple streaming questions with bounded memory~\cite{lyu2026pivot}. Existing memory is generally a single shared resource managed by one agent or pipeline, so local evidence and orchestrator-reviewed shared memory are not separate learning targets. BiFocal instead assigns detailed observations to Readers and exposes a chronological global timeline with source times and stored-frame handles, while learning policies for both roles.

\section{Method}
\label{sec:method}

\begin{figure*}[t]
\centering
\includegraphics[width=\textwidth]{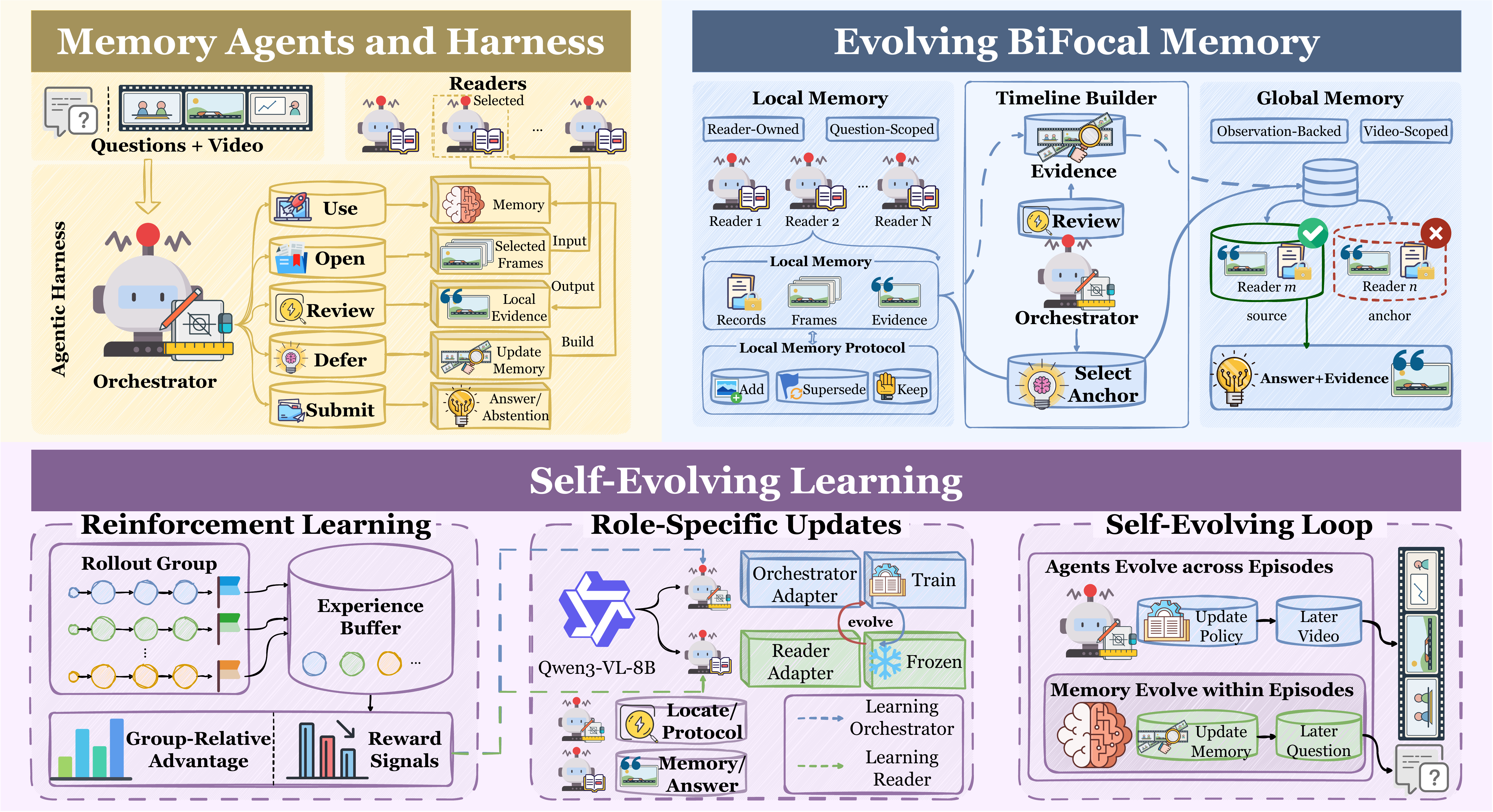}
\caption{\textbf{BiFocal.} \emph{Top left:} O selects one action at a time, and the harness validates and executes it. Readers $R_1,\ldots,R_N$ own separate video branches and return local records. \emph{Top right:} local records feed a chronological global timeline whose entries keep source time, revision, and a stored-frame handle. \emph{Bottom:} O and the Reader role train with separate adapters over one base model, one role at a time, across validated rounds (see Section~\ref{sec:learn}).}
\label{fig:system}
\end{figure*}

BiFocal separates memory by scope. Each Reader owns detailed observations from the branches it has inspected, while the team shares a chronological timeline of representative stored observations. Figure~\ref{fig:system} shows the system.

\subsection{Problem Formulation}
\label{sec:prelim}

A long video $V$ of duration $D$ is cut once into $N=\lceil D/64\rceil$ chunks of 64 seconds. A question $q$ supplies visible options $\mathcal{O}=\{o_1,\dots,o_m\}$ and is paired with a hidden correct option $y$. A read of chunk $c_k$ returns at most $F_k=32$ selected frames and a Reader report. An episode may spend at most $C_{\max}$ model calls and $F_{\max}$ frames (values in Section~\ref{sec:exp-setting}).

The system answers under a matched budget. Let $\hat y_\pi(q)$ be the answer produced by policy $\pi$. The task objective is
\begin{equation}
\begin{aligned}
\pi^{\star}={}&\arg\max_{\pi}\ \mathbb{E}\left[\mathbf{1}\{\hat y_\pi(q)=y\}\right]\\
&\text{s.t.}\ \ C(\pi)\le C_{\max},\; F(\pi)\le F_{\max}.
\end{aligned}
\label{eq:task}
\end{equation}
When a benchmark provides an evidence window $[t^{\ast},t^{\ast}{+}\Delta]$, it is used only for evaluation and diagnostics. It never appears in a policy-visible state.

The proposed system uses two learned roles inside a harness-managed environment. The orchestrator \textbf{O} holds the question and options, selects where to look and which protocol action to take, and writes the final answer. The Reader agents $\{R_i\}$ are assigned to branches $\{B_i\}$, inspect only their own branches, update their local memory, and return answer-bearing reports. All Readers share one Reader policy and adapter but own separate local records. O never sees a frame it did not request, and Readers never write directly into global memory, which the harness builds from the stored local observations of all Readers.

\subsection{BiFocal Memory Agents}
\label{sec:memory}

\textbf{Local memory.} A local record is written by one Reader for one question and one branch. It is
\begin{equation}
m=(id,\ \mathit{scope},\ \tau,\ \phi,\ w,\ \hat y,\ z,\ \kappa,\ s,\ v),
\label{eq:record}
\end{equation}
where $\tau$ is the time span, $\phi$ the stored frames, $w$ the Reader's words, $\hat y$ its branch answer, $z$ a supporting quote, $\kappa$ the call and Reader provenance, $s$ a lifecycle status, and $v$ a version. The local scope contains the video, question, Reader, and branch. A Reader patch chooses one memory operation: \textsc{add} creates a record, \textsc{supersede} writes a new version linked to the event it corrects, and \textsc{keep} commits no change. The harness rejects any patch with unresolved frames, invalid parents, out-of-branch spans, duplicate keys, or token cost above the limit.

Local records are append-only. A supersession preserves the superseded version, and \textsc{keep} leaves memory unchanged. This makes every Reader description auditable even when it is later replaced or rejected.

\textbf{Global memory.} Global memory is a video-scoped chronological ledger built from stored local observations. Memory persists across the questions of a video, so items written for an earlier question can serve later ones. The harness assigns observations to 128-second chapters by start time and keeps one representative observation per chapter, up to 12 displayed entries. Each entry retains its exact source time, compact summary, source revision, and stored-frame handle. Local and global entries may share a source revision. At retrieval, the harness selects the entry that best matches the current question and reconstructs the displayed claim from that entry's source, so \textsc{review} and \textsc{submit} remain tied to the selected revision. The ledger does not synthesize causal links or copy prior answer letters.

\textbf{Protocol.} O acts one turn at a time through the typed interface
\textsc{use}~$R_{i,j}$ or \textsc{defer}~$R_{i,j}$ for an available memory item,
\textsc{open}~$B_i$, \textsc{read}~$B_i$, \textsc{review}~$R_{i,j}$,
\textsc{submit}~$\hat y$ with a cited revision and quote, or \textsc{abstain}.
\textsc{open} exposes branch metadata and prior revisions. \textsc{read} invokes
the selected branch Reader. \textsc{review} checks an answer-bearing revision
against its stored frames and quote; it tests grounding, not answer correctness.
\textsc{use} opens and checks an eligible earlier memory item, while \textsc{defer}
continues without it. \textsc{submit} names the source of the answer, and the
harness verifies that the source and quote were observed. An abstention is a valid
terminal but earns no task credit. Memory use is recorded in stages: an item is
\emph{exposed} when it appears in a role's memory list, then opened and checked, and finally \emph{cited} by \textsc{submit}
(definitions in Appendix~\ref{app:evaluation}).

\subsection{Self-Evolving Learning}
\label{sec:learn}

The harness is the training environment. A training unit is a \emph{video group}: two questions of one video, asked in order with memory persisting across them. For each video group the harness samples a \emph{rollout group} of four independent policy rollouts of that two-question sequence, all from the same initial memory. The harness executes legal actions, validates memory patches, updates versions, charges calls and frames, and returns observations. A sampled trajectory records the exact tokens and log probabilities produced by O and the Readers. Memory writes become visible immediately within a rollout, but an adapter is updated only after a batch of rollouts has been validated.

\textbf{Reward.} Both roles learn from the same task outcome and pay for their own resource use,
\begin{equation}
r_O=r_{\mathrm{task}}-c_O,\qquad r_R=r_{\mathrm{task}}-c_R.
\label{eq:reward}
\end{equation}
The task outcome $r_{\mathrm{task}}$ is $+1$ for a correct answer and negative for a wrong answer, an abstention, or an incomplete episode (values in Appendix~\ref{app:training}). A rollout's reward is the mean over its two questions. The cost $c_O$ is a capped penalty, at most about $0.10$, on the calls and frames that O requests. The cost $c_R$ is the same kind of penalty on the calls and frames that the Readers spend, and it applies while O is frozen. Evidence gain, grounding, memory reuse, progress, and protocol terms are logged as diagnostics and are not trained. O therefore learns where to inspect, when to use or defer global memory, and when to stop. The Readers learn what to keep and report.

\textbf{Group-relative updates.} For either role, let $r_i$ be the reward of rollout $i$ in its rollout group of $G=4$, and let $A_i=\mathrm{clip}\big((r_i-\bar r)/\sigma_r,-c_A,c_A\big)$ be its reward standardized within the group as in GRPO~\cite{shao2024deepseekmath}, with $c_A=2$ for O. Let $\pi_\theta$, $\pi_{\mathrm{old}}$, and $\pi_{\mathrm{ref}}$ denote the trainable adapter, the policy that sampled the rollouts, and the frozen reference, and let $\rho_{i,t}=\pi_\theta(a_{i,t})/\pi_{\mathrm{old}}(a_{i,t})$ be the importance ratio of sampled token $a_{i,t}$. The shared objective is clipped PPO~\cite{schulman2017ppo} with a KL penalty and an entropy bonus,
\begin{equation}
\begin{aligned}
\mathcal{J}(\theta)=\mathbb{E}\Bigg[&\frac{1}{T}\sum_{i,t}
\min\!\left(\rho_{i,t}A_i,\right.\\
&\left.\mathrm{clip}(\rho_{i,t},1{-}\epsilon,1{+}\epsilon)A_i\right)\\
&-\beta\,\mathrm{KL}(\pi_\theta\Vert\pi_{\mathrm{ref}})+\eta\,\mathcal{H}(\pi_\theta)\Bigg],
\end{aligned}
\label{eq:group-relative}
\end{equation}
where $T$ is the number of sampled tokens, $\mathcal{H}$ is the policy entropy, and $\epsilon,\beta,\eta$ are given in Appendix~\ref{app:training}. Training minimizes $-\mathcal{J}$. O updates its action and review tokens. The Reader updates its sampled memory-content and branch-answer tokens. O and the Reader roles use separate LoRA adapters over one frozen-vision base model, and all Reader agents share the Reader adapter. At each update stage, one role adapter is trainable while the other remains frozen. The candidate adapter is promoted only after it passes a paired validation check against the previous adapter and a clean episode audit before the next round begins.

\section{Experiments}
\label{sec:experiments}

\subsection{Experimental Setting}
\label{sec:exp-setting}

\paragraph{Benchmarks.}
We evaluate on \textbf{LVBench}~\cite{wang2024lvbench} (103 videos, 1,549 questions),
\textbf{Video-MME}~\cite{fu2024videomme} (900 videos, 2,700 questions), 
\textbf{LongVideoBench}~\cite{wu2024longvideobench} (752 videos, 1,337 questions),  and
\textbf{MLVU}~\cite{zhou2024mlvu} (test split, 509 questions on 344 videos). Videos
have several questions (about 15 on LVBench, 3 on Video-MME, 1.5 on MLVU test, and
1.8 on LongVideoBench), asked in released order with memory persisting across them. Q1 is a
video's first question and Q2+ every later one; we report this split for LVBench and
Video-MME. We train and validate on MLVU dev
and never train or select checkpoints on the test sets. LVBench,
Video-MME, and LongVideoBench therefore test zero-shot transfer, and MLVU test is
in-distribution (Appendix~\ref{app:evaluation}).

\paragraph{Baselines.}
Table~\ref{tab:main-results} lists three base models, Qwen3-VL~\cite{qwen2025qwen3vl},
GPT-5~\cite{openai2025gpt5}, and Gemini 2.5 Pro~\cite{geminiteam2025gemini25}, and 14 agent
baselines grouped as single-agent, multi-agent, and memory-based. Matched comparisons,
ablations, and analyses use only Qwen3-VL-8B arms that we run in the same harness with the
same items and caps (Sections~\ref{sec:budget-reuse} and~\ref{sec:ablation}); BiFocal at B2
is the Full BiFocal arm of those sections. The \emph{fixed prompt policy} is our matched
baseline: the same harness and memory as BiFocal, driven by hand-written fixed prompts with
no BiFocal training, and the initial policy of training. The \emph{no-memory} arm clears
memory before each question, the \emph{single-agent control} gives one agent the same
question, memory interface, and caps with no Reader/O split, and the
\emph{shuffled-reward control} trains like BiFocal with rewards permuted within each
rollout group.

\paragraph{Implementation details.}
All matched arms use Qwen3-VL-8B, and each read returns at most 32 frames at a long side of 768 pixels. Budgets B1, B2, and
B3 cap each question at 10, 20, and 40 calls and 128, 256, and 512 frames, identically
for every arm. Trained arms learn LoRA~\cite{hu2022lora} adapters over three
rounds on MLVU dev, each with 300 video groups of two questions and four rollouts per
group, updating O and then the Reader (Sections~\ref{sec:learn}, \ref{app:training}).

\paragraph{Evaluation metrics.}
The metric is each benchmark's official accuracy, never pooled across benchmarks;
wrong answers, abstentions, and protocol failures score zero. We also report calls,
frames, tokens, and time per question. Contrasts are paired
accuracy differences averaged over three seeds. An interval spans the lowest lower
and highest upper bound of the three per-seed 95\% intervals, clustered by video
(Table~\ref{tab:supp-seeds}). B2 is the main budget, and the primary
contrast is Full BiFocal against the fixed prompt policy.

\subsection{Main Results}
\label{sec:main-results}

\providecommand{\resultvenue}[1]{{\scriptsize\textcolor{gray}{[#1]}}}
\providecommand{\up}[1]{{\,\scriptsize\textcolor[HTML]{0B5394}{$\uparrow$#1}}}
\providecommand{\dn}[1]{{\,\scriptsize\textcolor[HTML]{A84300}{$\downarrow$#1}}}

\begin{table*}[!t]
\centering
\scriptsize
\setlength{\tabcolsep}{2pt}
\renewcommand{\arraystretch}{0.95}
\caption{\textbf{Long-video accuracy (\%).} \textbf{Frames} is the frame budget or sampling rate; for our two rows it is the mean actual frames per question across the four benchmarks (B2, three seeds). Both use Qwen3-VL-8B; the fixed prompt policy is BiFocal without training. $\uparrow$/$\downarrow$: gain or drop in accuracy points relative to Qwen3-VL-8B.}
\label{tab:main-results}
\begin{tabular}{@{}>{\raggedright\arraybackslash}m{0.205\textwidth}%
>{\raggedright\arraybackslash}m{\dimexpr 0.185\textwidth-2\tabcolsep\relax}%
>{\centering\arraybackslash}m{\dimexpr 0.105\textwidth-2\tabcolsep\relax}%
>{\centering\arraybackslash}m{\dimexpr 0.090\textwidth-2\tabcolsep\relax}%
>{\centering\arraybackslash}m{\dimexpr 0.090\textwidth-2\tabcolsep\relax}%
>{\centering\arraybackslash}m{\dimexpr 0.115\textwidth-2\tabcolsep\relax}%
>{\centering\arraybackslash}m{\dimexpr 0.1050\textwidth-2\tabcolsep\relax}%
>{\centering\arraybackslash}m{\dimexpr 0.1050\textwidth-2\tabcolsep\relax}@{}}
\toprule
\textbf{Method} & \textbf{Model} & \textbf{Frames} & \textbf{LVBench} &
\textbf{MLVU} & \textbf{LongVideoBench} & \multicolumn{2}{c}{\textbf{Video-MME (w/o sub.)}} \\
\cmidrule(lr){7-8}
& & & & & & \textbf{Overall} & \textbf{Long} \\
\midrule
\multicolumn{8}{@{}l}{\textit{Base Models}} \\
Qwen3-VL-8B~\cite{qwen2025qwen3vl}\,\resultvenue{TR'25} & Qwen3-VL-8B & 256 & 48.0 & 68.1 & 54.5 & 60.4 & 51.2 \\
GPT-5~\cite{openai2025gpt5}\,\resultvenue{System Card} & GPT-5 & 256 & 59.8 \up{11.8} & 78.3 \up{10.2} & 61.8 \up{7.3} & 77.3 \up{16.9} & 68.4 \up{17.2} \\
Gemini 2.5 Pro~\cite{geminiteam2025gemini25}\,\resultvenue{TR'25} & Gemini 2.5 Pro & 512 & 69.0 \up{21.0} & 81.2 \up{13.1} & 65.2 \up{10.7} & 80.6 \up{20.2} & 71.3 \up{20.1} \\
\midrule
\multicolumn{8}{@{}l}{\textit{Single-agent}} \\
VideoZoomer~\cite{ding2026videozoomer}\,\resultvenue{ICLR'26} & Qwen2.5-VL-7B & 128 & 41.5 \dn{6.5} & 68.8 \up{0.7} & 57.7 \up{3.2} & 65.2 \up{4.8} & 55.8 \up{4.6} \\
VTS~\cite{zhang2026vts}\,\resultvenue{NeurIPS'26} & Qwen3-VL-8B & 274 & 54.7 \up{6.7} & 58.2 \dn{9.9} & 60.2 \up{5.7} & 67.5 \up{7.1} & 58.3 \up{7.1} \\
MambaMia~\cite{kim2026mambamia}\,\resultvenue{AAAI'26} & Qwen2-7B & 384 & 44.6 \dn{3.4} & 68.0 \dn{0.1} & 57.1 \up{2.6} & 58.3 \dn{2.1} & 50.3 \dn{0.9} \\
VideoChat-Flash~\cite{li2026videochatflash}\,\resultvenue{ICLR'26} & Qwen2-7B & 64--512 & 48.2 \up{0.2} & 74.7 \up{6.6} & 64.7 \up{10.2} & 65.3 \up{4.9} & 55.4 \up{4.2} \\
MSJoE~\cite{tan2026msjoe}\,\resultvenue{CVPR'26} & Qwen2.5-VL-7B & 64 & 51.1 \up{3.1} & 75.1 \up{7.0} & 62.2 \up{7.7} & 66.2 \up{5.8} & 57.4 \up{6.2} \\
\midrule
\multicolumn{8}{@{}l}{\textit{Multi-agent}} \\
VideoSEAL~\cite{qiu2026videoseal}\,\resultvenue{ICML'26} & Qwen3-8B + Qwen2.5-VL-7B & 64 & 55.1 \up{7.1} & 68.2 \up{0.1} & 62.0 \up{7.5} & 62.9 \up{2.5} & 53.4 \up{2.2} \\
VideoMind~\cite{liu2026videomind}\,\resultvenue{ICLR'26} & Qwen2-VL-7B & 32--150 & 40.8 \dn{7.2} & 64.4 \dn{3.7} & 59.1 \up{4.6} & 58.2 \dn{2.2} & 49.2 \dn{2.0} \\
SVAgent~\cite{yang2026svagent}\,\resultvenue{CVPR'26} & Qwen3-VL-8B & 8 & 40.8 \dn{7.2} & 65.6 \dn{2.5} & 61.0 \up{6.5} & 63.1 \up{2.7} & 53.9 \up{2.7} \\
ReAgent-V~\cite{zhou2025reagentv}\,\resultvenue{NeurIPS'25} & Qwen2.5-VL-7B & 35 & 40.5 \dn{7.5} & 60.7 \dn{7.4} & 58.7 \up{4.2} & 60.7 \up{0.3} & 49.8 \dn{1.4} \\
\midrule
\multicolumn{8}{@{}l}{\textit{Memory-based}} \\
VideoLucy~\cite{zuo2025videolucy}\,\resultvenue{NeurIPS'25} & Qwen2.5-VL-7B + R1 & 1--2fps & 58.8 \up{10.8} & 76.1 \up{8.0} & 60.2 \up{5.7} & 72.5 \up{12.1} & 66.8 \up{15.6} \\
QViC-MF~\cite{yamao2026qvic}\,\resultvenue{CVPR'26} & LLaVA-Video-7B-Qwen2 & 2fps & 50.2 \up{2.2} & 59.4 \dn{8.7} & 57.3 \up{2.8} & 63.4 \up{3.0} & 54.0 \up{2.8} \\
FluxMem~\cite{xie2026fluxmem}\,\resultvenue{CVPR'26} & Qwen2.5-VL-7B & 1fps & 48.2 \up{0.2} & 73.1 \up{5.0} & 61.1 \up{6.6} & 65.3 \up{4.9} & 54.0 \up{2.8} \\
WorldMM~\cite{yeo2026worldmm}\,\resultvenue{CVPR'26} & Qwen3-VL-8B + GPT-5-mini & 1fps & 55.4 \up{7.4} & 74.3 \up{6.2} & 60.2 \up{5.7} & 66.1 \up{5.7} & 66.0 \up{14.8} \\
MemStream~\cite{agarwal2026memstream}\,\resultvenue{NeurIPS'26} & Qwen2.5-VL-7B & 128 & 48.1 \up{0.1} & 66.3 \dn{1.8} & 57.9 \up{3.4} & 62.3 \up{1.9} & 54.2 \up{3.0} \\
\midrule
\rowcolor{bifocallightblue}
\textit{BiFocal (Ours)} & & & & & & & \\
\rowcolor{bifocallightblue}
Fixed prompt policy & Qwen3-VL-8B & 206 & 56.2 \up{8.2} & 69.4 \up{1.3} & 61.9 \up{7.4} & 66.2 \up{5.8} & 63.6 \up{12.4} \\
\rowcolor{bifocallightblue}
\textbf{BiFocal} & Qwen3-VL-8B & 160 & \textbf{59.4} \up{11.4} & \textbf{77.2} \up{9.1} & \textbf{64.9} \up{10.4} & \textbf{73.2} \up{12.8} & \textbf{68.9} \up{17.7} \\
\bottomrule
\end{tabular}
\end{table*}

\paragraph{Comparison with baselines.}
Table~\ref{tab:main-results} compares BiFocal with relevant baselines. With an 8B
backbone and at most 32 frames per read, BiFocal scores above Qwen3-VL-8B and above
every listed agent system on all four benchmarks, by 0.2 to 1.1 points over the best. Some of those systems add a separate language model (Model column), and GPT-5 and
Gemini 2.5 Pro still lead on some benchmarks. Relative to Qwen3-VL-8B, the fixed prompt
policy (the same harness without training) adds +8.2, +5.8, +1.3, and +7.4 points on
LVBench, Video-MME, MLVU test, and LongVideoBench, and training adds a further +3.2, +7.0,
+7.8, and +3.0. The harness thus accounts for most of the lead on LVBench and
LongVideoBench, and training for more on Video-MME and MLVU test.

\subsection{Budget and Cross-Question Reuse}
\label{sec:budget-reuse}

\begin{figure*}[!t]
\centering
\includegraphics[width=\textwidth]{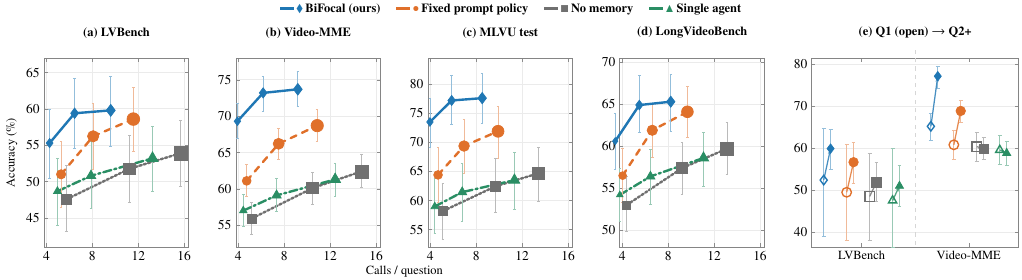}
\caption{\textbf{Budget and cross-question reuse.} (a)--(d) Accuracy versus calls per question; lines run from B1 to B3, marker area is mean frames. (e) Q1 (open) and Q2+ (filled) accuracy at B2. Mean over three seeds; bars: 95\% interval envelope (Section~\ref{sec:exp-setting}).}
\label{fig:budget-reuse}
\end{figure*}

\begin{figure*}[!t]
\centering
\includegraphics[width=\textwidth]{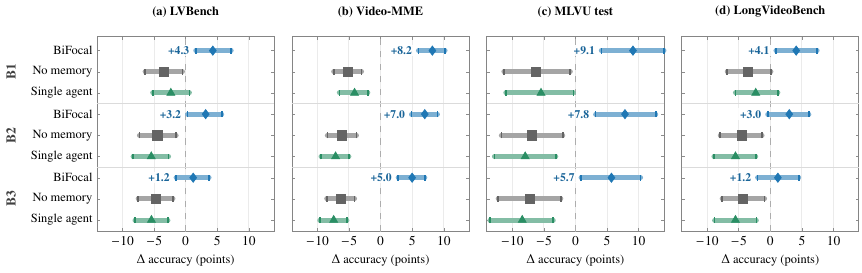}
\caption{\textbf{Matched-budget differences.} Accuracy difference from the fixed prompt policy (points); bars: 95\% interval envelope.}
\label{fig:matched-budget-effects}
\end{figure*}

\paragraph{Better answers from fewer reads.}
Figure~\ref{fig:budget-reuse}(a)--(d) plots accuracy against the calls each arm spends. In
every cell, Full BiFocal sits above and to the left of the other arms: it is more accurate
with the fewest calls, 14 to 20\% fewer than the fixed prompt policy, and also uses fewer
frames, 14 to 21\% fewer tokens, and 14 to 24\% less wall time (Table~\ref{tab:supp-budget}).
Extra budget does not close the gap: at B3, twice the budget, the fixed prompt policy is
still 0.8 to 5.3 points less accurate than Full BiFocal at B2, using 56 to 59\% of its calls
and 51 to 54\% of its frames. At B1, Full BiFocal uses 51 to 57\% of the fixed prompt
policy's B2 calls and is more accurate on Video-MME and MLVU test (+3.1, +4.1) but lower on
LVBench ($-$0.9) and LongVideoBench ($-$1.3).
\paragraph{The advantage is largest when the budget is tight.}
Figure~\ref{fig:matched-budget-effects} gives the paired differences from the fixed prompt
policy. Full BiFocal is ahead in all twelve cells, and the interval excludes zero in nine
(Table~\ref{tab:supp-seeds}). The largest gains are on MLVU test at B1 (+9.1, [+4.1,
+14.0]) and Video-MME at B1 (+8.2, [+6.0, +10.2]); at the main budget B2 they are +7.8 and
+7.0. On every benchmark the gain shrinks as the cap grows, for example from +8.2 to +5.0 on
Video-MME, because from B2 to B3 the fixed prompt policy gains 2.2 to 2.5 points but Full
BiFocal only 0.4 to 0.5 (Table~\ref{tab:supp-budget}), so BiFocal saturates earlier. This
fits reuse mattering most when reads are scarce, since a loose cap lets the
fixed prompt policy re-read what BiFocal retrieves from memory. LVBench and LongVideoBench gain less (+1.2 to +4.3).

\paragraph{Memory and the Reader/O split both contribute.}
Relative to the fixed prompt policy, removing memory lowers accuracy in all twelve cells,
by 3.4 to 7.2 points, and costs 9 to 39\% more calls, so memory saves reads as well as
improving answers. Removing the Reader/O split also lowers accuracy in every cell, by 2.3
to 8.4 points, while the single agent spends 0.9 to 1.1 times the calls of the fixed prompt
policy, so the loss is not from reading less, and the benefit is not from memory alone.

\paragraph{Memory pays off on later questions.}
Figure~\ref{fig:budget-reuse}(e) separates a video's first question from later ones. A
first question has nothing to reuse. On Video-MME, removing memory leaves first questions
almost unchanged ($-$0.4 points) and costs $-$9.0 points on later ones, and the no-memory
arm scores the same on both groups (60.4 and 59.9), so later questions are not simply
easier. On LVBench later questions are somewhat easier even without memory (48.5 against
52.0), and the effect of removing memory is smaller ($-$1.0 against $-$4.7; only 103 first
questions). Two matched episodes (Table~\ref{tab:supp-examples}, Appendix~\ref{app:failure}) illustrate this.

\subsection{Ablation Study}
\label{sec:ablation}

\begin{table*}[!t]
\centering
\scriptsize
\setlength{\tabcolsep}{1.97pt}
\renewcommand{\arraystretch}{1.05}
\caption{\textbf{Ablations at B2.} \emph{(a)} Each row alters one component of the fixed prompt policy. \emph{(b)} Learning arms share items, caps, and rollouts; shared, O-only, and Reader-only use double LoRA rank (equal parameters and FLOPs); the shuffled-reward control permutes rewards within rollout groups. Accuracy (\%) is mean$\pm$std over three seeds. $\Delta$ is the paired difference from the fixed prompt policy (points, 95\% interval envelope, Section~\ref{sec:exp-setting}), bold if it excludes zero; the last row compares BiFocal with the shared adapter. Cost columns are LVBench means; \textit{Unsup.}: share of cited items failing the post-hoc checker (--: not measured). Position splits: Table~\ref{tab:supp-ablation-strata}.}
\label{tab:ablation-memory}
\label{tab:ablation-learning}
\begin{tabular}{lccccccccccc}
\toprule
 & \multicolumn{2}{c}{\textbf{LVBench}} & \multicolumn{2}{c}{\textbf{Video-MME}} & \multicolumn{2}{c}{\textbf{MLVU test}} & \multicolumn{2}{c}{\textbf{LongVideoBench}} & \multicolumn{3}{c}{\textbf{LVBench cost}} \\
\cmidrule(lr){2-3}\cmidrule(lr){4-5}\cmidrule(lr){6-7}\cmidrule(lr){8-9}\cmidrule(lr){10-12}
\textbf{Variant} & \textbf{Acc.} & \textbf{$\Delta$ [95\% CI]} & \textbf{Acc.} & \textbf{$\Delta$ [95\% CI]} & \textbf{Acc.} & \textbf{$\Delta$ [95\% CI]} & \textbf{Acc.} & \textbf{$\Delta$ [95\% CI]} & \textbf{Calls/Q} & \textbf{Frames/Q} & \textbf{Unsup.} \\
\midrule
\multicolumn{12}{l}{\textit{(a) Memory ablations}} \\
No memory & 51.8$\pm$0.2 & \textbf{-4.5 [-7.3, -1.5]} & 60.1$\pm$0.1 & \textbf{-6.1 [-8.4, -3.8]} & 62.4$\pm$0.1 & \textbf{-6.9 [-11.7, -2.0]} & 57.4$\pm$0.1 & \textbf{-4.5 [-7.9, -1.3]} & 11.24 & 254.8 & 0.026 \\
Unrelated memory & 54.0$\pm$0.3 & -2.2 [-5.2, +0.6] & 63.4$\pm$0.4 & \textbf{-2.8 [-5.1, -0.4]} & 66.1$\pm$0.6 & -3.3 [-8.7, +2.1] & 59.5$\pm$0.4 & -2.4 [-6.0, +0.9] & 8.57 & 239.4 & 0.094 \\
Flat memory & 54.5$\pm$0.3 & -1.7 [-4.6, +1.1] & 64.0$\pm$0.5 & -2.2 [-4.8, +0.3] & 66.9$\pm$0.4 & -2.4 [-7.7, +2.9] & 60.1$\pm$0.5 & -1.8 [-5.6, +1.7] & 8.48 & 236.1 & 0.048 \\
No grounding check & 55.1$\pm$0.5 & -1.1 [-4.1, +1.7] & 65.0$\pm$0.4 & -1.2 [-3.7, +1.3] & 67.6$\pm$0.3 & -1.8 [-7.2, +3.1] & 60.8$\pm$0.5 & -1.1 [-4.9, +2.3] & 7.76 & 229.7 & 0.187 \\
No revisions & 55.5$\pm$0.3 & -0.7 [-3.7, +2.2] & 65.4$\pm$0.2 & -0.8 [-3.1, +1.5] & 68.2$\pm$0.2 & -1.2 [-6.1, +3.5] & 61.2$\pm$0.5 & -0.7 [-4.5, +3.0] & 7.98 & 233.0 & 0.112 \\
\rowcolor{gray!12}
Fixed prompt policy & 56.2$\pm$0.2 & ref. & 66.2$\pm$0.1 & ref. & 69.4$\pm$0.2 & ref. & 61.9$\pm$0.1 & ref. & 8.07 & 237.1 & 0.031 \\
\midrule
\multicolumn{12}{l}{\textit{(b) Learning ablations}} \\
Fixed prompt policy & 56.2$\pm$0.2 & ref. & 66.2$\pm$0.1 & ref. & 69.4$\pm$0.2 & ref. & 61.9$\pm$0.1 & ref. & 8.07 & 237.1 & -- \\
Shuffled-reward control & 55.9$\pm$0.5 & -0.3 [-3.3, +2.5] & 65.8$\pm$0.2 & -0.4 [-2.5, +1.7] & 68.9$\pm$0.1 & -0.4 [-4.9, +4.1] & 61.6$\pm$0.5 & -0.3 [-3.8, +3.3] & 8.11 & 237.8 & -- \\
O-only & 56.9$\pm$0.4 & +0.7 [-2.7, +3.8] & 67.5$\pm$0.4 & +1.3 [-1.2, +3.6] & 70.8$\pm$0.2 & +1.4 [-3.5, +6.1] & 62.7$\pm$0.2 & +0.8 [-2.7, +4.2] & 7.73 & 225.1 & -- \\
Reader-only & 57.4$\pm$0.5 & +1.2 [-1.9, +4.4] & 68.4$\pm$0.4 & +2.2 [-0.2, +4.7] & 71.8$\pm$0.4 & +2.4 [-2.5, +7.2] & 63.2$\pm$0.4 & +1.3 [-2.4, +4.9] & 7.47 & 215.9 & -- \\
Shared adapter & 57.7$\pm$0.2 & +1.5 [-1.5, +4.3] & 68.8$\pm$0.4 & \textbf{+2.6 [+0.5, +5.0]} & 72.2$\pm$0.5 & +2.9 [-2.7, +7.8] & 63.5$\pm$0.5 & +1.6 [-2.0, +5.1] & 7.32 & 210.7 & -- \\
\rowcolor{bifocallightblue}
BiFocal (Ours) & 59.4$\pm$0.6 & \textbf{+3.2 [+0.3, +5.7]} & 73.2$\pm$0.5 & \textbf{+7.0 [+4.9, +9.0]} & 77.2$\pm$0.5 & \textbf{+7.8 [+3.2, +12.7]} & 64.9$\pm$0.5 & +3.0 [-0.4, +6.1] & 6.46 & 180.2 & -- \\
\addlinespace[1pt]
BiFocal $-$ shared adapter &  & +1.7 [-1.5, +4.6] &  & \textbf{+4.4 [+2.0, +7.0]} &  & +5.0 [-0.7, +10.3] &  & +1.4 [-2.1, +5.4] &  &  &  \\
\bottomrule
\end{tabular}
\end{table*}

\paragraph{Local and global memory both matter.}
Table~\ref{tab:ablation-memory}(a) changes one part of the memory at a time:
\textit{no memory} clears it before each question, \textit{unrelated memory} inserts memory
from another video, \textit{flat memory} merges local and global records, \textit{no
grounding check} skips the comparison with stored frames, and \textit{no revisions}
disables \textsc{supersede}. Every change lowers mean accuracy on all four benchmarks, in
the same order on each: no memory (4.5 to 6.9 points), unrelated memory (2.2 to 3.3),
flat memory (1.7 to 2.4), no grounding check (1.1 to 1.8), and no revisions (0.7 to 1.2).
Only removing memory has an interval below zero on every benchmark, so the benefit of
memory is clear. Memory from the wrong video removes about half of it, so the benefit
depends on what the memory contains, and merging the two scopes costs 1.7 to 2.4 points.
Skipping the grounding check costs little accuracy but raises the share of unsupported
citations from 0.031 to 0.187, so the check mainly protects citation quality. That share
also rises without revisions (0.112) and with unrelated memory (0.094), which suggests that
stale or foreign entries are cited when nothing screens them out. Calls change little
except without memory (7.8 to 8.6 against 8.07 for the fixed prompt policy on LVBench, and
11.24 without memory), so these accuracy gaps are not bought with extra reads. Of the 16
partial-ablation intervals only one (unrelated memory on Video-MME) excludes zero, so we
read these effects as trends, not as established differences between the variants.

\paragraph{Role-specific policies make learning pay off.}
Training on the true reward raises mean accuracy for every trained variant, and separate
role policies gain the most. One shared adapter for both roles gains 1.5 to 2.9 points, and
separate O and Reader adapters add a further 1.7, 4.4, 5.0, and 1.4 points on the four
benchmarks; this direct contrast excludes zero on Video-MME only. Training only one role
gains less than the shared adapter (0.7 to 1.4 points for O-only, 1.2 to 2.4 for
Reader-only), and the four trained arms rank in the same order on every benchmark: O-only,
Reader-only, shared adapter, BiFocal. The two single-role gains do not add up to the full gain: on every benchmark
BiFocal's gain exceeds their sum (for example +7.8 against +3.8 on MLVU test), which
suggests the roles are complementary, although the single-role intervals include zero.
O's choices about where to look and what to cite depend on what the Readers wrote, and the
Readers' notes help only if O uses them, so separate adapters let each role specialize
while one adapter must trade the two objectives off. The shuffled-reward control stays
within 0.4 points of the fixed prompt policy, so the gain comes from the reward and not
from extra updates.

\paragraph{Self-evolving memory.}
Table~\ref{tab:evolution} follows the validation block over three rounds. Full BiFocal
improves from 69.6 at the initial checkpoint to 73.8, 76.2, and 77.9 after rounds 1 to 3.
Each round passes the keep rule, but the gains over the parent shrink (+4.1, +2.4, and
+1.7 points), so about half of the total improvement arrives in round 1. Memory use changes
with accuracy: cited items per question rise from 0.34 to 1.51 and duplicate writes fall
from 1.26 to 0.25, so the team writes less redundant memory and relies on it more. The
other arms reach 70.9 to 72.7 at round three, and the shuffled-reward control shows neither
the accuracy gain nor the memory change (69.8), tying both to the reward.

\begin{table}[t]
\centering
\scriptsize
\setlength{\tabcolsep}{4.74pt}
\renewcommand{\arraystretch}{1.08}
\caption{\textbf{Self-evolution on the MLVU dev validation block.} \emph{(a)} Accuracy per round (0: fixed prompt policy); bold: best arm. \emph{(b)} Cited and duplicate items per question. Mean$\pm$std over three seeds; the block also selects checkpoints. Intervals: Table~\ref{tab:supp-evolution}.}
\label{tab:evolution}
\begin{tabular}{lcccc}
\toprule
\textbf{Round} & \textbf{0} & \textbf{1} & \textbf{2} & \textbf{3} \\
\midrule
\multicolumn{5}{l}{\textit{(a) Validation accuracy (\%)}} \\
\rowcolor{bifocallightblue}
BiFocal (Ours) & 69.6$\pm$0.2 & \textbf{73.8$\pm$0.2} & \textbf{76.2$\pm$0.3} & \textbf{77.9$\pm$0.3} \\
Shared adapter & 69.6$\pm$0.2 & 70.8$\pm$0.2 & 72.1$\pm$0.1 & 72.7$\pm$0.1 \\
Reader-only & 69.6$\pm$0.2 & 70.4$\pm$0.2 & 71.4$\pm$0.3 & 72.4$\pm$0.2 \\
O-only & 69.6$\pm$0.2 & 69.9$\pm$0.2 & 70.6$\pm$0.2 & 70.9$\pm$0.3 \\
Shuffled-reward control & 69.6$\pm$0.2 & 69.9$\pm$0.3 & 69.6$\pm$0.1 & 69.8$\pm$0.3 \\
\midrule
\multicolumn{5}{l}{\textit{(b) Memory items per question}} \\
\rowcolor{bifocallightblue}
BiFocal, cited & 0.34$\pm$0.02 & 0.70$\pm$0.02 & 1.11$\pm$0.03 & 1.51$\pm$0.06 \\
\rowcolor{bifocallightblue}
BiFocal, duplicates & 1.26$\pm$0.08 & 0.95$\pm$0.05 & 0.59$\pm$0.02 & 0.25$\pm$0.01 \\
Shuffled, cited & 0.34$\pm$0.02 & 0.36$\pm$0.02 & 0.37$\pm$0.02 & 0.36$\pm$0.02 \\
Shuffled, duplicates & 1.26$\pm$0.08 & 1.27$\pm$0.07 & 1.35$\pm$0.04 & 1.26$\pm$0.06 \\
\bottomrule
\end{tabular}
\end{table}

\subsection{Further Analysis}
\label{sec:further-analysis}

\begin{figure}[t]
\centering
\includegraphics[width=\columnwidth]{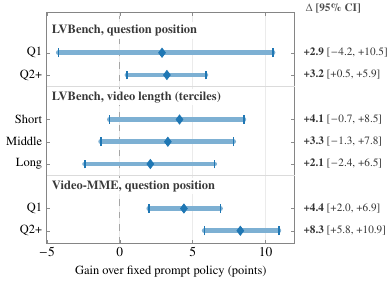}
\caption{\textbf{Where the gain appears.} Full BiFocal minus the fixed prompt policy at B2; 3-seed mean, bars: 95\% interval envelope.}
\label{fig:stratified-effects}
\end{figure}

\paragraph{Where the gain appears.}
Figure~\ref{fig:stratified-effects} splits the gain by question position. A first
question has nothing to reuse, so its gain on Video-MME (+4.4) reflects better reading and
allocation alone, and the nearly twice larger gain on later questions (+8.3) adds the
benefit of reuse. Reuse also shows in effort: on later questions Full BiFocal makes 17.6
to 19.9\% fewer calls and about 19 to 22\% fewer frames than on first questions
(Table~\ref{tab:analysis-strata}, Appendix~\ref{app:evaluation}), while the no-memory arm
makes about the same number of calls on both. From first to later questions, accuracy rises by 12.0 points for Full BiFocal, 8.1 for the
fixed prompt policy, and $-$0.5 without memory on Video-MME, and by 7.5, 7.2, and 3.5 on
LVBench (Figure~\ref{fig:budget-reuse}(e)), so training adds to reuse mainly on Video-MME. On LVBench the gains on first and later
questions are similar (+2.9 and +3.2), but the first-question interval is wide, so reuse
is one source of the gain and not the only one. By video length the gain falls from +4.1 to +2.1 as calls rise, but every interval includes zero.

\paragraph{Memory use.}
Table~\ref{tab:supp-memory-use} (Appendix~\ref{app:evaluation}) follows memory items from
exposure to citation on LVBench later questions. Full BiFocal opens 72.3\% of exposed items,
checks 62.3\%, and cites 49.9\%, against 66.0\%, 55.2\%, and 40.4\% for the fixed prompt
policy, so it uses more of the memory it is shown, also on first questions (43.0\% against 37.4\% cited). O starts 42.6\% of its cited items against 28.7\%, so training raises O's own use of
memory, not only the harness's. Memory use gets cleaner: episodes citing stale or
out-of-scope memory fall from 59 to 22 and unsupported citations from 33 to 17
(Table~\ref{tab:supp-failure}, Appendix~\ref{app:failure}). Full BiFocal answers 920
LVBench questions correctly against 871, and the gain comes with fewer memory errors.

\paragraph{Transfer and stability.}
The gain appears on all four test sets at every budget though trained only on MLVU dev.
It is largest on MLVU test, the in-distribution set, and on Video-MME (+7.8 and +7.0 at
B2) and smaller but positive on LVBench and LongVideoBench (+3.2 and +3.0). All 36 seed-level
gains are positive, std at most 0.6 (Tables~\ref{tab:supp-budget}, \ref{tab:supp-seeds}).

\section{Conclusion}
\label{sec:conclusion}

We presented BiFocal, a multi-agent system for long-video reasoning with local Reader
memory, a shared global timeline, and separate policies for O and the Readers. Under
matched budgets it is more accurate than the fixed prompt policy on four
benchmarks in all twelve settings, with 14 to 20\% fewer calls, and removing memory,
merging scopes, or sharing an adapter lowers mean accuracy.

\section*{LLM Usage Statement}
\label{sec:llm-usage}

We used large language models for two purposes: refining and polishing the
wording of the manuscript, and assisting with coding, using Codex and Claude Code. All other content was checked
by the authors, who take responsibility for the text, code, and reported results.

{
    \small
    \bibliographystyle{ieeenat_fullname}
    \bibliography{main}
}

\clearpage
\appendix

\section{Evaluation Protocol and Analysis Details}
\label{app:evaluation}

\subsection{Splits, denominators, and exclusions}
Training and validation use MLVU dev. We allocate dev at the video level by task
type and duration into a training pool and a validation block. MLVU dev and MLVU
test are separate splits of the official release, so neither pool overlaps a test set.
The \emph{training} pool trains the adapters. The \emph{validation} block is
scored after every round and decides whether that round's update is kept.
LVBench, Video-MME, LongVideoBench, and MLVU test are test sets. The harness, prompts, budgets,
and arms are fixed before each test run, and earlier diagnostic runs are excluded.
Memory resets at the start of each video and persists across its questions. Each
test set is scored with the checkpoint fixed beforehand, and supplies the test-set
accuracy columns in the budget, ablation, and further-analysis tables. Every arm
is run with three seeds. For a trained arm, a seed is an independent training run
with its own rollout seeds, followed by one greedy test evaluation of each trained
adapter. For an untrained arm, the three seeds are three test runs. Reported values are means over seeds, and
each interval spans the lowest lower bound and the highest upper bound of the three
per-seed intervals (Table~\ref{tab:supp-seeds}).
Validation accuracies are reported separately. The validation block is never used as test data
(Table~\ref{tab:supp-splits}). An attempted task episode has valid infrastructure
and ends either with a valid terminal or a protocol failure. Terminal failure
rates use attempted task episodes as the denominator, so protocol failures
remain visible. Accuracy also uses attempted episodes: wrong answers, abstentions,
and protocol failures receive zero credit. A completed clean episode reaches a
valid terminal without harness recovery, infrastructure failure, or a cut run;
clean status is an audit diagnostic and does not change the accuracy denominator.
Excluded infrastructure and cut episodes are counted separately and never pooled
into a reported arm. Infrastructure failures were re-run, so $N$ is the full
question count; the failed attempts are excluded from $N$. For cross-question
analyses, Q1 denotes the first question in a video and Q2+ denotes later questions.
Rates for Q1 and Q2+ use their own denominators.

Table~\ref{tab:supp-splits} lists the benchmark and split counts. MLVU dev and MLVU test are
disjoint splits of the official release. MLVU test has 509 questions on 344
videos, about 1.5 per video, and its Q1/Q2+ split is not reported.

\begin{table*}[!t]
\centering
\scriptsize
\setlength{\tabcolsep}{3pt}
\renewcommand{\arraystretch}{0.92}
\caption{\textbf{Split and denominator inventory.} MLVU dev is allocated by task type and duration into training and validation pools; dev and test are separate splits of the official release. MLVU test has 509 questions on 344 videos. Test sets are never used for training or checkpoint selection.}
\label{tab:supp-splits}
\begin{tabular}{@{}>{\raggedright\arraybackslash}m{0.16\textwidth}%
>{\raggedright\arraybackslash}m{\dimexpr 0.16\textwidth-2\tabcolsep\relax}%
>{\raggedright\arraybackslash}m{\dimexpr 0.30\textwidth-2\tabcolsep\relax}%
>{\centering\arraybackslash}m{\dimexpr 0.14\textwidth-2\tabcolsep\relax}%
>{\centering\arraybackslash}m{\dimexpr 0.11\textwidth-2\tabcolsep\relax}%
>{\centering\arraybackslash}m{\dimexpr 0.13\textwidth-2\tabcolsep\relax}@{}}
\toprule
\textbf{Pool} & \textbf{Source} & \textbf{Use} & \textbf{Share of dev} &
\textbf{Videos} & \textbf{Questions} \\
\midrule
Training & MLVU dev & Adapter training, three rounds of 600 questions & 70\% & 935 & 1,817 \\
Validation & MLVU dev & Scored each round; keeps or rolls back & 30\% & 399 & 776 \\
\midrule
MLVU test & MLVU test & In-distribution test & n/a & 344 & 509 \\
LVBench & LVBench & Zero-shot transfer test & n/a & 103 & 1,549 \\
Video-MME & Video-MME & Zero-shot transfer test & n/a & 900 & 2,700 \\
LongVideoBench & LongVideoBench & Zero-shot transfer, validation & n/a & 752 & 1,337 \\
\bottomrule
\end{tabular}
\end{table*}

\subsection{Training volume}
MLVU dev has 2,593 questions on 1,334 videos, or 1.94 questions per video.
The split gives 935 training videos (1,817 questions) and a
validation block of 399 videos (776 questions). Training
proceeds in three rounds. A round has 300 video groups, each two questions of one
video, or 600 questions. The three rounds therefore use 900 video groups and 1,800
questions, on 900 of the 935 training videos and 1,800 of the 1,817 training
questions. Each round uses a disjoint slice of the training pool, so no training
question repeats. The unit is a video group with its ordered pair of questions,
because memory persists across them. For each video group the harness samples a
rollout group of four independent policy rollouts of the pair, and every trained
arm sees the same video groups in the same order. The shuffled-reward control runs
the same rollouts, learning rate, and compute, with rewards permuted within each
rollout group. After each round,
the validation block checks whether the candidate update is safe to keep. With
discordance $d$ between two policies on $n$ questions, the paired 95\%
half-width is about $1.96\sqrt{d/n}$ before clustering. At $d=0.2$ and $n=776$
the conservative pre-run bound is 3.1 points. The observed per-seed
parent-checkpoint intervals have a half-width of about 1.6 points
(Table~\ref{tab:supp-evolution} shows their envelope), consistent with $d\approx0.05$. Keeping an update therefore uses a predeclared non-regression rule. An
\emph{accepted} disposition in Table~\ref{tab:supp-evolution} records that outcome. Because the same block selects updates
and draws the trend, the test sets carry all accuracy claims.

Before round one, a short calibration run (one
arm, about 100 training questions, four rollouts per group) measures seconds per episode, the share
of groups with identical rewards, and the discordance $d$ between the calibration
adapter and the initial policy on the validation block. That adapter is
discarded. The run sets the keep margin, which is then fixed before round one.

\subsection{Training setup and compute}
\label{app:training}
The supplementary code lists every setting; we give the ones needed to read the results.

\textbf{Budgets.} Every arm has the same per-question caps. B1, B2, and B3 allow 10,
20, and 40 calls; 128, 256, and 512 frames; 60, 180, and 420 seconds; and 16k, 32k,
and 64k tokens. Gold answers and evidence annotations never enter a policy-visible
state.

\textbf{Adapters and optimizer.} Each role has a BF16 LoRA adapter on the query,
key, value, and output projections with the vision encoder frozen (rank 16, alpha
32, dropout 0.05; 15.3M trainable parameters per role). The shared, O-only, and Reader-only ablation adapters use rank 32. We use AdamW with learning
rate $5{\times}10^{-7}$ for O and $1{\times}10^{-6}$ for the Reader, weight decay 0.01, gradient clipping 1.0, no scheduler, and
two epochs per phase. The PPO ratio clip is $\epsilon=0.2$, the KL weight is
$\beta=0.001$, and the entropy weight is $\eta=0.01$. Rollouts are generated with
vLLM replicas that load the current adapters (BF16, 512-token completion limit).

\textbf{Reward.} Both roles are rewarded by task outcome minus cost. A correct
answer scores $+1$. A wrong answer, an abstention with accepted evidence, a protocol
abstention, and an incomplete episode score $-1$, and an abstention without accepted
evidence scores $-0.5$. A capped call and frame cost of at most about $0.10$ is
subtracted: O's cost for O's calls and frames, and the Reader's cost for the Reader's
calls and frames, with O frozen. A rollout's reward is the mean over its two
questions. Rewards are standardized within each rollout group of four, and O's
standardized advantage is clipped to $[-2,2]$. Earlier low-reward turns remain in the training data; only infrastructure-corrupted transitions are excluded.

\textbf{Schedule.} Each round has 300 video groups (600 questions) and four rollouts
per group. Phase A updates O with the Reader frozen, and Phase B updates the Reader
with the new O frozen, using fresh rollouts. Four video groups form one optimizer
step, so a phase has 75 steps per epoch and 150 over two epochs. A round has 300
steps and three rounds have 900, 450 per role. One step processes $4\times4\times2=32$
rollout-questions. One training seed produces 14,400 role-rollouts (3 rounds
$\times$ 2 phases $\times$ 300 video groups $\times$ 4 rollouts $\times$ 2 questions).

\textbf{Compute.} One seed needs about 169 accelerator-hours of teacher-forced
training and 258 of rollout collection, or 427 in total. The three seeds need about
1,281 accelerator-hours on RTX 3090 GPUs, about 337 hours on four cards at 95\%
end-to-end utilization (about 14 days). Excluding the vision encoder, one seed costs
about $3.28{\times}10^{19}$ training-plus-replay FLOPs and $1.46{\times}10^{19}$
collection-inference FLOPs, so the three seeds cost about $1.42{\times}10^{20}$
FLOPs for training and rollout collection. Training seeds use distinct rollout seeds.

\subsection{Matched-budget analysis}
Within each budget level, Full BiFocal, Fixed prompt policy, No memory,
and Single agent use the same items, prompt revision,
question order, decoding settings, call cap, frame cap, time cap, and new-token
cap. The caps increase across budget levels. We report actual calls, frames,
tokens, and wall-clock time rather than capped values. Paired differences use
95 percent confidence intervals, and a difference counts as clear when the whole
interval is on one side of zero.
The budget levels are fixed on MLVU dev before any test run, and the same levels
apply to all four benchmarks.

Table~\ref{tab:supp-budget} reports every arm, budget, and benchmark: 1,549
questions from 103 videos on LVBench, 2,700 from 900 on Video-MME, 509 on
MLVU test, and 1,337 from 752 on LongVideoBench. Intervals are clustered by video (103, 900, 344, and 752 videos). Full BiFocal
has higher accuracy and fewer calls and frames than the fixed prompt policy in every
cell. Its combined interval excludes zero in nine of the twelve cells. The three cells
with an interval that includes zero are LVBench at B3, LongVideoBench at B2, and
LongVideoBench at B3.
Table~\ref{tab:supp-seeds} lists the per-seed contrasts. The gain is positive in all
36 seed-by-cell contrasts, and its lower bound is above zero in 29 of them (all but
LVBench at B3, LongVideoBench at B3, and one seed of LongVideoBench at B2). On every
benchmark the three seeds agree within 1.5 points at every budget, so the reported
gains are stable under retraining.

\subsection{Paired differences and per-seed contrasts}
Tables~\ref{tab:ablation-memory} and~\ref{tab:supp-ablation-strata} report benchmark
accuracy, paired differences, and other measures for the ablation arms. Table~\ref{tab:supp-budget} gives paired
differences for the four budget arms on all four benchmarks, with the fixed prompt
policy as the reference, and Table~\ref{tab:supp-seeds} gives the per-seed contrasts. Full BiFocal versus the fixed prompt policy and versus the
shared adapter are the primary learning contrasts.

\begin{table*}[!t]
\centering
\scriptsize
\setlength{\tabcolsep}{3pt}
\renewcommand{\arraystretch}{0.92}
\caption{\textbf{Matched-budget results.} Within each budget, all arms share items, prompts, order, decoding, and caps. Values are means over three seeds; $\pm$ is the standard deviation across seeds. Full BiFocal uses the last accepted checkpoint; the other arms use the fixed prompt policy (reference). $\Delta$ is the paired difference from the reference with a 95\% CI, clustered by video. Calls, frames, tokens, and wall time are per-question means; abstain and protocol failure are percentages of attempted episodes. Bold: best of accuracy, calls, frames, tokens, and wall time within each budget.}
\label{tab:supp-budget}
\begin{tabular}{@{}>{\centering\arraybackslash}m{0.0600\textwidth}%
>{\raggedright\arraybackslash}m{\dimexpr 0.1500\textwidth-2\tabcolsep\relax}%
>{\centering\arraybackslash}m{\dimexpr 0.0950\textwidth-2\tabcolsep\relax}%
>{\centering\arraybackslash}m{\dimexpr 0.1250\textwidth-2\tabcolsep\relax}%
>{\centering\arraybackslash}m{\dimexpr 0.0680\textwidth-2\tabcolsep\relax}%
>{\centering\arraybackslash}m{\dimexpr 0.0780\textwidth-2\tabcolsep\relax}%
>{\centering\arraybackslash}m{\dimexpr 0.0780\textwidth-2\tabcolsep\relax}%
>{\centering\arraybackslash}m{\dimexpr 0.0660\textwidth-2\tabcolsep\relax}%
>{\centering\arraybackslash}m{\dimexpr 0.0780\textwidth-2\tabcolsep\relax}%
>{\centering\arraybackslash}m{\dimexpr 0.0780\textwidth-2\tabcolsep\relax}%
>{\centering\arraybackslash}m{\dimexpr 0.1130\textwidth-2\tabcolsep\relax}@{}}
\toprule
\textbf{Budget} & \textbf{Arm} & \textbf{Acc.} & \textbf{$\Delta$ [95\% CI]} &
\textbf{Calls/Q} & \textbf{Frames/Q} & \textbf{Tokens/Q} & \textbf{Wall/Q} &
\textbf{Abstain} & \textbf{Protocol fail} & \textbf{Pairs / clusters} \\
\midrule
\multicolumn{11}{@{}l}{\textit{LVBench}} \\
\rowcolor{bifocallightblue}
B1 & \textbf{Full BiFocal} & \textbf{55.3$\pm$0.5} & \textbf{+4.3 [+1.6, +7.2]} & \textbf{4.28} & \textbf{115.7} & \textbf{9,476} & \textbf{6.15} & 1.44 & 0.79 & 1,549 / 103 \\
B1 & Fixed prompt policy & 51.0$\pm$0.1 & ref. & 5.28 & 125.1 & 11,951 & 8.11 & 1.70 & 1.10 & 1,549 / 103 \\
B1 & No memory & 47.6$\pm$0.2 & -3.4 [-6.4, -0.5] & 5.75 & 127.4 & 13,927 & 9.01 & 2.17 & 1.89 & 1,549 / 103 \\
B1 & Single agent & 48.7$\pm$0.2 & -2.3 [-5.2, +0.6] & 4.96 & 121.3 & 10,430 & 6.70 & 1.89 & 1.36 & 1,549 / 103 \\
\cmidrule(lr){1-11}
\rowcolor{bifocallightblue}
B2 & \textbf{Full BiFocal} & \textbf{59.4$\pm$0.6} & \textbf{+3.2 [+0.3, +5.7]} & \textbf{6.46} & \textbf{180.2} & \textbf{14,506} & \textbf{9.12} & 0.69 & 0.39 & 1,549 / 103 \\
B2 & Fixed prompt policy & 56.2$\pm$0.2 & ref. & 8.07 & 237.1 & 17,819 & 11.15 & 0.73 & 0.32 & 1,549 / 103 \\
B2 & No memory & 51.8$\pm$0.2 & -4.5 [-7.3, -1.5] & 11.24 & 254.8 & 26,813 & 16.46 & 1.21 & 0.92 & 1,549 / 103 \\
B2 & Single agent & 50.8$\pm$0.1 & -5.4 [-8.3, -2.7] & 7.88 & 223.7 & 17,005 & 10.64 & 0.92 & 0.54 & 1,549 / 103 \\
\cmidrule(lr){1-11}
\rowcolor{bifocallightblue}
B3 & \textbf{Full BiFocal} & \textbf{59.8$\pm$0.5} & \textbf{+1.2 [-1.5, +3.7]} & \textbf{9.60} & \textbf{284.6} & \textbf{21,021} & \textbf{12.72} & 0.54 & 0.26 & 1,549 / 103 \\
B3 & Fixed prompt policy & 58.6$\pm$0.1 & ref. & 11.57 & 351.4 & 26,333 & 16.05 & 0.58 & 0.34 & 1,549 / 103 \\
B3 & No memory & 53.9$\pm$0.1 & -4.7 [-7.5, -2.0] & 15.68 & 414.7 & 38,369 & 22.23 & 0.90 & 0.54 & 1,549 / 103 \\
B3 & Single agent & 53.2$\pm$0.1 & -5.4 [-8.0, -2.8] & 13.24 & 383.0 & 27,764 & 16.55 & 0.69 & 0.39 & 1,549 / 103 \\
\midrule
\multicolumn{11}{@{}l}{\textit{Video-MME}} \\
\rowcolor{bifocallightblue}
B1 & \textbf{Full BiFocal} & \textbf{69.3$\pm$0.5} & \textbf{+8.2 [+6.0, +10.2]} & \textbf{3.92} & \textbf{104.3} & \textbf{8,422} & \textbf{5.46} & 1.28 & 0.84 & 2,700 / 900 \\
B1 & Fixed prompt policy & 61.1$\pm$0.1 & ref. & 4.73 & 110.3 & 10,359 & 7.02 & 1.56 & 1.03 & 2,700 / 900 \\
B1 & No memory & 55.9$\pm$0.1 & -5.2 [-7.4, -3.0] & 5.14 & 112.4 & 12,071 & 7.79 & 2.08 & 1.80 & 2,700 / 900 \\
B1 & Single agent & 57.0$\pm$0.1 & -4.1 [-6.5, -2.0] & 4.43 & 107.0 & 9,039 & 5.79 & 1.84 & 1.16 & 2,700 / 900 \\
\cmidrule(lr){1-11}
\rowcolor{bifocallightblue}
B2 & \textbf{Full BiFocal} & \textbf{73.2$\pm$0.5} & \textbf{+7.0 [+4.9, +9.0]} & \textbf{6.16} & \textbf{167.5} & \textbf{13,415} & \textbf{8.41} & 0.68 & 0.29 & 2,700 / 900 \\
B2 & Fixed prompt policy & 66.2$\pm$0.1 & ref. & 7.51 & 213.4 & 16,075 & 10.04 & 0.81 & 0.39 & 2,700 / 900 \\
B2 & No memory & 60.1$\pm$0.1 & -6.1 [-8.4, -3.8] & 10.45 & 230.2 & 24,209 & 14.81 & 1.14 & 0.94 & 2,700 / 900 \\
B2 & Single agent & 59.1$\pm$0.1 & -7.1 [-9.4, -5.0] & 7.33 & 201.3 & 15,340 & 9.58 & 0.88 & 0.52 & 2,700 / 900 \\
\cmidrule(lr){1-11}
\rowcolor{bifocallightblue}
B3 & \textbf{Full BiFocal} & \textbf{73.7$\pm$0.5} & \textbf{+5.0 [+2.8, +7.0]} & \textbf{9.18} & \textbf{261.9} & \textbf{19,501} & \textbf{11.77} & 0.53 & 0.22 & 2,700 / 900 \\
B3 & Fixed prompt policy & 68.7$\pm$0.1 & ref. & 10.87 & 319.5 & 23,993 & 14.59 & 0.62 & 0.33 & 2,700 / 900 \\
B3 & No memory & 62.4$\pm$0.1 & -6.3 [-8.5, -4.1] & 14.73 & 376.9 & 34,959 & 20.21 & 0.86 & 0.58 & 2,700 / 900 \\
B3 & Single agent & 61.3$\pm$0.1 & -7.4 [-9.6, -5.3] & 12.43 & 348.1 & 25,296 & 15.05 & 0.70 & 0.42 & 2,700 / 900 \\
\midrule
\multicolumn{11}{@{}l}{\textit{MLVU}} \\
\rowcolor{bifocallightblue}
B1 & \textbf{Full BiFocal} & \textbf{73.5$\pm$0.5} & \textbf{+9.1 [+4.1, +14.0]} & \textbf{3.97} & \textbf{102.9} & \textbf{9,063} & \textbf{5.59} & 1.44 & 1.11 & 509 / 344 \\
B1 & Fixed prompt policy & 64.4$\pm$0.2 & ref. & 4.68 & 107.8 & 10,898 & 7.01 & 1.57 & 1.38 & 509 / 344 \\
B1 & No memory & 58.2$\pm$0.2 & -6.2 [-11.3, -0.8] & 5.09 & 109.8 & 12,700 & 7.80 & 2.29 & 2.23 & 509 / 344 \\
B1 & Single agent & 59.0$\pm$0.3 & -5.4 [-11.0, -0.3] & 4.39 & 105.0 & 9,510 & 5.79 & 2.29 & 1.38 & 509 / 344 \\
\cmidrule(lr){1-11}
\rowcolor{bifocallightblue}
B2 & \textbf{Full BiFocal} & \textbf{77.2$\pm$0.5} & \textbf{+7.8 [+3.2, +12.7]} & \textbf{5.86} & \textbf{155.5} & \textbf{13,580} & \textbf{8.09} & 0.85 & 0.52 & 509 / 344 \\
B2 & Fixed prompt policy & 69.4$\pm$0.2 & ref. & 6.94 & 194.4 & 15,784 & 9.37 & 0.92 & 0.39 & 509 / 344 \\
B2 & No memory & 62.4$\pm$0.1 & -6.9 [-11.7, -2.0] & 9.67 & 209.9 & 23,756 & 13.83 & 1.51 & 0.85 & 509 / 344 \\
B2 & Single agent & 61.4$\pm$0.1 & -7.9 [-12.8, -3.1] & 6.78 & 183.4 & 15,063 & 8.94 & 0.92 & 0.33 & 509 / 344 \\
\cmidrule(lr){1-11}
\rowcolor{bifocallightblue}
B3 & \textbf{Full BiFocal} & \textbf{77.6$\pm$0.5} & \textbf{+5.7 [+0.9, +10.3]} & \textbf{8.52} & \textbf{237.9} & \textbf{19,202} & \textbf{11.02} & 0.52 & 0.26 & 509 / 344 \\
B3 & Fixed prompt policy & 71.9$\pm$0.2 & ref. & 9.90 & 286.7 & 23,209 & 13.41 & 0.65 & 0.46 & 509 / 344 \\
B3 & No memory & 64.7$\pm$0.3 & -7.2 [-12.2, -2.3] & 13.41 & 338.4 & 33,817 & 18.58 & 0.85 & 0.79 & 509 / 344 \\
B3 & Single agent & 63.5$\pm$0.2 & -8.4 [-13.5, -3.5] & 11.33 & 312.5 & 24,470 & 13.83 & 0.72 & 0.26 & 509 / 344 \\
\midrule
\multicolumn{11}{@{}l}{\textit{LongVideoBench}} \\
\rowcolor{bifocallightblue}
B1 & \textbf{Full BiFocal} & \textbf{60.6$\pm$0.5} & \textbf{+4.1 [+0.9, +7.4]} & \textbf{3.39} & \textbf{83.9} & \textbf{7,395} & \textbf{4.64} & 1.47 & 0.82 & 1,337 / 752 \\
B1 & Fixed prompt policy & 56.5$\pm$0.1 & ref. & 4.04 & 89.2 & 8,795 & 5.89 & 1.73 & 1.14 & 1,337 / 752 \\
B1 & No memory & 53.0$\pm$0.1 & -3.5 [-6.8, +0.1] & 4.39 & 90.8 & 9,576 & 6.54 & 2.20 & 1.96 & 1,337 / 752 \\
B1 & Single agent & 54.2$\pm$0.1 & -2.3 [-5.5, +1.2] & 3.79 & 86.5 & 8,251 & 4.87 & 1.99 & 1.38 & 1,337 / 752 \\
\cmidrule(lr){1-11}
\rowcolor{bifocallightblue}
B2 & \textbf{Full BiFocal} & \textbf{64.9$\pm$0.5} & \textbf{+3.0 [-0.4, +6.1]} & \textbf{5.49} & \textbf{137.9} & \textbf{11,973} & \textbf{7.38} & 0.73 & 0.26 & 1,337 / 752 \\
B2 & Fixed prompt policy & 61.9$\pm$0.1 & ref. & 6.62 & 177.9 & 14,423 & 8.70 & 0.70 & 0.41 & 1,337 / 752 \\
B2 & No memory & 57.4$\pm$0.1 & -4.5 [-7.9, -1.3] & 9.22 & 192.1 & 20,089 & 12.84 & 1.32 & 0.85 & 1,337 / 752 \\
B2 & Single agent & 56.4$\pm$0.1 & -5.5 [-8.9, -2.3] & 6.46 & 167.9 & 14,084 & 8.30 & 0.88 & 0.59 & 1,337 / 752 \\
\cmidrule(lr){1-11}
\rowcolor{bifocallightblue}
B3 & \textbf{Full BiFocal} & \textbf{65.3$\pm$0.5} & \textbf{+1.2 [-2.0, +4.5]} & \textbf{8.23} & \textbf{219.2} & \textbf{17,931} & \textbf{10.36} & 0.47 & 0.15 & 1,337 / 752 \\
B3 & Fixed prompt policy & 64.1$\pm$0.1 & ref. & 9.68 & 269.0 & 21,093 & 12.77 & 0.62 & 0.35 & 1,337 / 752 \\
B3 & No memory & 59.7$\pm$0.1 & -4.4 [-7.6, -0.9] & 13.12 & 317.4 & 28,584 & 17.69 & 0.94 & 0.62 & 1,337 / 752 \\
B3 & Single agent & 58.6$\pm$0.1 & -5.5 [-8.8, -2.2] & 11.08 & 293.2 & 24,136 & 13.17 & 0.62 & 0.38 & 1,337 / 752 \\
\bottomrule
\end{tabular}
\end{table*}

\begin{table*}[!t]
\centering
\scriptsize
\setlength{\tabcolsep}{3pt}
\renewcommand{\arraystretch}{0.92}
\caption{\textbf{Per-seed gain of Full BiFocal} over the fixed prompt policy, in points with 95\% CI, for three independent seeds. \textit{Combined} is the mean with the lowest lower and highest upper bound, as in Table~\ref{tab:supp-budget}. Bold: lower bound above zero.}
\label{tab:supp-seeds}
\begin{tabular}{@{}llcccc@{}}
\toprule
\textbf{Benchmark} & \textbf{Budget} & \textbf{Seed 0} & \textbf{Seed 1} & \textbf{Seed 2} & \textbf{Combined} \\
\midrule
LVBench & B1 & \textbf{+3.6 [+1.6, +6.0]} & \textbf{+4.6 [+2.2, +7.0]} & \textbf{+4.7 [+2.3, +7.2]} & \textbf{+4.3 [+1.6, +7.2]} \\
 & B2 & \textbf{+2.6 [+0.3, +5.0]} & \textbf{+3.1 [+0.7, +5.5]} & \textbf{+3.8 [+1.4, +5.7]} & \textbf{+3.2 [+0.3, +5.7]} \\
 & B3 & +0.7 [-1.5, +3.1] & +1.6 [-0.8, +3.7] & +1.3 [-1.1, +3.7] & +1.2 [-1.5, +3.7] \\
\midrule
Video-MME & B1 & \textbf{+7.8 [+6.0, +9.5]} & \textbf{+8.7 [+7.0, +10.2]} & \textbf{+8.1 [+6.4, +9.8]} & \textbf{+8.2 [+6.0, +10.2]} \\
 & B2 & \textbf{+6.4 [+4.9, +8.1]} & \textbf{+7.1 [+5.4, +8.8]} & \textbf{+7.5 [+5.8, +9.0]} & \textbf{+7.0 [+4.9, +9.0]} \\
 & B3 & \textbf{+4.9 [+3.2, +6.6]} & \textbf{+4.6 [+2.8, +6.3]} & \textbf{+5.5 [+3.8, +7.0]} & \textbf{+5.0 [+2.8, +7.0]} \\
\midrule
MLVU & B1 & \textbf{+9.1 [+4.9, +13.3]} & \textbf{+8.7 [+4.1, +12.9]} & \textbf{+9.5 [+5.3, +14.0]} & \textbf{+9.1 [+4.1, +14.0]} \\
 & B2 & \textbf{+7.6 [+3.4, +11.8]} & \textbf{+7.3 [+3.2, +11.5]} & \textbf{+8.6 [+4.4, +12.7]} & \textbf{+7.8 [+3.2, +12.7]} \\
 & B3 & \textbf{+5.9 [+1.7, +10.1]} & \textbf{+4.9 [+0.9, +9.1]} & \textbf{+6.3 [+2.1, +10.3]} & \textbf{+5.7 [+0.9, +10.3]} \\
\midrule
LongVideoBench & B1 & \textbf{+4.3 [+1.4, +7.2]} & \textbf{+3.4 [+0.9, +6.3]} & \textbf{+4.6 [+1.7, +7.4]} & \textbf{+4.1 [+0.9, +7.4]} \\
 & B2 & +2.6 [-0.4, +5.5] & \textbf{+3.5 [+0.6, +6.1]} & \textbf{+2.9 [+0.1, +5.8]} & +3.0 [-0.4, +6.1] \\
 & B3 & +1.4 [-1.5, +4.3] & +0.5 [-2.0, +3.4] & +1.7 [-1.2, +4.5] & +1.2 [-2.0, +4.5] \\
\bottomrule
\end{tabular}
\end{table*}

\begin{table}[!t]
\centering
\scriptsize
\setlength{\tabcolsep}{3pt}
\renewcommand{\arraystretch}{0.92}
\caption{\textbf{Gain by question position and video length.} Full BiFocal minus the fixed prompt policy at B2, in points, with 95\% CI clustered by video. Calls and frames are per-question means for Full BiFocal.}
\label{tab:analysis-strata}
\begin{tabular}{@{}>{\raggedright\arraybackslash}m{0.3291\columnwidth}%
>{\centering\arraybackslash}m{\dimexpr 0.1139\columnwidth-2\tabcolsep\relax}%
>{\centering\arraybackslash}m{\dimexpr 0.2532\columnwidth-2\tabcolsep\relax}%
>{\centering\arraybackslash}m{\dimexpr 0.1392\columnwidth-2\tabcolsep\relax}%
>{\centering\arraybackslash}m{\dimexpr 0.1646\columnwidth-2\tabcolsep\relax}@{}}
\toprule
\textbf{Stratum} & \textbf{Pairs} & \textbf{$\Delta$ [95\% CI]} &
\textbf{Calls/Q} & \textbf{Frames/Q} \\
\midrule
\multicolumn{5}{@{}l}{\textit{LVBench, question position}} \\
Q1 & 103 & +2.9 [-4.2, +10.5] & 7.73 & 225.9 \\
Q2+ & 1,446 & +3.2 [+0.5, +5.9] & 6.37 & 177.0 \\
\midrule
\multicolumn{5}{@{}l}{\textit{LVBench, video length (terciles)}} \\
Shortest & 517 & +4.1 [-0.7, +8.5] & 5.26 & 146.1 \\
Middle & 516 & +3.3 [-1.3, +7.8] & 6.42 & 177.3 \\
Longest & 516 & +2.1 [-2.4, +6.5] & 7.69 & 217.2 \\
\midrule
\multicolumn{5}{@{}l}{\textit{Video-MME, question position}} \\
Q1 & 900 & +4.4 [+2.0, +6.9] & 7.10 & 191.6 \\
Q2+ & 1,800 & +8.3 [+5.8, +10.9] & 5.69 & 155.5 \\
\bottomrule
\end{tabular}
\end{table}

\begin{table*}[!t]
\centering
\scriptsize
\setlength{\tabcolsep}{4pt}
\renewcommand{\arraystretch}{1.05}
\caption{\textbf{Memory ablations by question position and video length at B2.} Same arms and conventions as Table~\ref{tab:ablation-memory}. One LVBench Q1 question moves accuracy by 0.97 points, so Q1 values take only a few distinct levels.}
\label{tab:supp-ablation-strata}
\begin{tabular}{@{}lcccccc@{}}
\toprule
 & \multicolumn{2}{c}{\textbf{LVBench Q1 (103 questions)}} & \multicolumn{2}{c}{\textbf{LVBench Q2+ (1,446)}} & \multicolumn{2}{c}{\textbf{Video-MME Long}} \\
\cmidrule(lr){2-3}\cmidrule(lr){4-5}\cmidrule(lr){6-7}
\textbf{Variant} & \textbf{Acc.} & \textbf{$\Delta$ [95\% CI]} & \textbf{Acc.} & \textbf{$\Delta$ [95\% CI]} & \textbf{Acc.} & \textbf{$\Delta$ [95\% CI]} \\
\midrule
No memory & 48.5$\pm$2.0 & -1.0 [-8.5, +7.2] & 52.0$\pm$0.2 & \textbf{-4.7 [-7.5, -1.7]} & 56.1$\pm$0.4 & \textbf{-7.5 [-11.7, -3.8]} \\
Unrelated memory & 49.5$\pm$2.6 & +0.0 [-7.5, +7.5] & 54.3$\pm$0.2 & -2.4 [-5.4, +0.6] & 59.4$\pm$0.5 & \textbf{-4.2 [-8.4, -0.1]} \\
Flat memory & 48.5$\pm$1.7 & -1.0 [-8.5, +6.4] & 54.9$\pm$0.4 & -1.8 [-5.0, +1.4] & 60.2$\pm$0.5 & \textbf{-3.4 [-6.9, -0.2]} \\
No grounding check & 48.5$\pm$2.6 & -1.0 [-8.8, +6.4] & 55.6$\pm$0.6 & -1.1 [-4.7, +2.2] & 61.3$\pm$0.4 & -2.3 [-6.1, +1.3] \\
No revisions & 49.5$\pm$2.0 & +0.0 [-7.9, +8.2] & 55.9$\pm$0.3 & -0.8 [-3.9, +2.5] & 62.0$\pm$0.5 & -1.6 [-4.7, +1.5] \\
\rowcolor{gray!12}
Fixed prompt policy & 49.5$\pm$2.6 & ref. & 56.7$\pm$0.3 & ref. & 63.6$\pm$0.5 & ref. \\
\bottomrule
\end{tabular}
\end{table*}

\subsection{Memory-use funnel}
An item is \emph{exposed} when it appears in a role's visible memory list. It is
\emph{mentioned} when the role names it, \emph{opened} when selected for use,
and \emph{checked} when compared with its stored frames. The checker then
records \emph{accepted} or \emph{rejected}. A submitted answer \emph{cites} an
item only when it names the checked source revision and quote. Every stage rate
uses exposed items for that arm and question position as its denominator. Q1
and Q2+ are reported separately. Each opened, checked, and cited step records
whether O or the harness started it. Harness review is real system memory use,
but it is not evidence that a learned policy chose memory. Exposure alone is not
evidence of memory use. Table~\ref{tab:supp-memory-use} gives item-level rates
over exposed items.

\begin{table*}[!t]
\centering
\scriptsize
\setlength{\tabcolsep}{1.5pt}
\renewcommand{\arraystretch}{1.0}
\caption{\textbf{LVBench memory-use funnel at B2.} Each stage is $n$ (\%) of exposed items ($N$), with the number started by O in brackets; the rest were harness steps, which are not evidence that a learned policy chose memory. Q1 items come from the current question, Q2+ items from earlier ones. Accepted and rejected partition checked items; cited items are a subset of checked. Bold: highest rate per position, excluding rejected.}
\label{tab:supp-memory-use}
\begin{tabular}{@{}>{\raggedright\arraybackslash}p{0.125\textwidth}>{\centering\arraybackslash}p{0.04\textwidth}>{\centering\arraybackslash}p{0.05\textwidth}>{\centering\arraybackslash}p{0.108\textwidth}>{\centering\arraybackslash}p{0.108\textwidth}>{\centering\arraybackslash}p{0.108\textwidth}>{\centering\arraybackslash}p{0.108\textwidth}>{\centering\arraybackslash}p{0.108\textwidth}>{\centering\arraybackslash}p{0.108\textwidth}@{}}
\toprule
\textbf{Arm} & \textbf{Pos.} & \textbf{Exp. $N$} & \textbf{Mentioned} & \textbf{Opened} & \textbf{Checked} & \textbf{Accepted} & \textbf{Rejected} & \textbf{Cited} \\
\midrule
Fixed prompt policy & Q1 & 350 & 291 (83.1) [78] & 230 (65.7) [64] & 194 (55.4) [52] & 165 (47.1) [44] & 29 (8.3) [8] & 131 (37.4) [35] \\
Fixed prompt policy & Q2+ & 4,916 & 4,188 (85.2) [1,214] & 3,245 (66.0) [961] & 2,714 (55.2) [793] & 2,321 (47.2) [681] & 393 (8.0) [112] & 1,986 (40.4) [570] \\
\addlinespace[2pt]
O-only & Q1 & 254 & 215 (84.6) [89] & 170 (66.9) [71] & 143 (56.3) [58] & 124 (48.8) [50] & 19 (7.5) [8] & 106 (41.7) [43] \\
O-only & Q2+ & 3,572 & 3,065 (85.8) [1,271] & 2,468 (69.1) [992] & 2,072 (58.0) [821] & 1,811 (50.7) [707] & 261 (7.3) [114] & 1,547 (43.3) [604] \\
\addlinespace[2pt]
Reader-only & Q1 & 302 & 258 (85.4) [107] & \textbf{203 (67.2) [84]} & 170 (56.3) [70] & \textbf{148 (49.0) [59]} & 22 (7.3) [11] & 123 (40.7) [49] \\
Reader-only & Q2+ & 4,237 & 3,652 (86.2) [1,512] & 2,924 (69.0) [1,191] & 2,441 (57.6) [981] & 2,123 (50.1) [843] & 318 (7.5) [138] & 1,830 (43.2) [722] \\
\addlinespace[2pt]
Full BiFocal & Q1 & 200 & \textbf{171 (85.5) [59]} & 134 (67.0) [48] & \textbf{114 (57.0) [41]} & \textbf{98 (49.0) [36]} & 16 (8.0) [5] & \textbf{86 (43.0) [31]} \\
Full BiFocal & Q2+ & 2,805 & \textbf{2,463 (87.8) [1,032]} & \textbf{2,028 (72.3) [864]} & \textbf{1,748 (62.3) [742]} & \textbf{1,529 (54.5) [648]} & 219 (7.8) [94] & \textbf{1,400 (49.9) [596]} \\
\bottomrule
\end{tabular}
\end{table*}

The funnel shows that training changes what the team does with memory, not only how
much it writes. On LVBench Q2+, where earlier-question memory is available, Full
BiFocal opens 72.3\% of exposed items against 66.0\% for the fixed prompt policy,
checks 62.3\% against 55.2\%, and cites 49.9\% against 40.4\%. The extra items it
checks are accepted: the accepted share rises from 47.2\% to 54.5\% while the rejected
share stays near 8\%. Full BiFocal reaches these rates from a smaller exposed set
(2,805 items against 4,916), consistent with the drop in duplicate writes in
Section~\ref{sec:ablation}, so more of what is shown to the team is worth using.

The roles also change who acts. O starts 42.6\% of Full BiFocal's cited Q2+ items
(596 of 1,400), against 28.7\% for the fixed prompt policy (570 of 1,986), and 42.6\% of its
opened items against 29.6\%. Training raises O's own use of memory, in addition to
the harness's. The two single-role arms each reach about 43\% cited on Q2+, and the
two-role system reaches 49.9\%, so the gains from training O and the Reader combine.
Q1 shows the same ordering, with Full BiFocal citing 43.0\% of exposed items against
37.4\%, so the effect does not depend on memory from earlier questions alone.

\subsection{Self-evolution protocol}
After round $r$, each arm's candidate is scored on the validation block, which is
video-disjoint from training, with fixed seeds, matched training compute, and a
paired interval against the parent checkpoint. The update is marked accepted,
rejected, or rolled back. A rollback returns to the parent checkpoint, and round
$r+1$ trains from it on its own slice. Every trained arm, including the
shuffled-reward control, is scored on the same block under the same keep rule,
and the checkpoint is fixed before the block is scored. The checkpoint reported
on the test sets is the last accepted one. The shuffled-reward control shows how far an update with no reward
signal moves the policy and how often the keep rule accepts it.

\begin{table}[!t]
\centering
\scriptsize
\setlength{\tabcolsep}{2pt}
\renewcommand{\arraystretch}{0.92}
\caption{\textbf{Validation rounds on MLVU dev.} All rows use the same 776-question block, which also selects the kept updates; test results are in Section~\ref{sec:budget-reuse}. Mean$\pm$std over three seeds; $\Delta$ is the paired difference from the parent checkpoint, with the per-seed interval envelope. Cited/Q is cited memory items per question. Arms follow Table~\ref{tab:ablation-learning}. Bold: best in each round.}
\label{tab:supp-evolution}
\begin{tabular}{@{}clccrl@{}}
\toprule
\textbf{Round} & \textbf{Arm} & \textbf{Acc.} & \textbf{$\Delta$ [95\% CI]} & \textbf{Cited/Q} & \textbf{Disposition} \\
\midrule
0 & Fixed prompt policy & 69.6$\pm$0.2 & ref. & 0.34$\pm$0.02 & Initial \\
1 & Shared adapter & 70.8$\pm$0.2 & +1.1 [-0.8, +2.9] & 0.57$\pm$0.03 & Accepted \\
2 & Shared adapter & 72.1$\pm$0.1 & +1.3 [-0.4, +3.4] & 0.78$\pm$0.03 & Accepted \\
3 & Shared adapter & 72.7$\pm$0.1 & +0.7 [-1.3, +2.4] & 0.91$\pm$0.04 & Accepted \\
1 & O-only & 69.9$\pm$0.2 & +0.3 [-1.4, +1.8] & 0.48$\pm$0.01 & Accepted \\
2 & O-only & 70.6$\pm$0.2 & +0.7 [-1.4, +2.5] & 0.62$\pm$0.04 & Accepted \\
3 & O-only & 70.9$\pm$0.3 & +0.3 [-1.6, +2.2] & 0.66$\pm$0.03 & Accepted \\
1 & Reader-only & 70.4$\pm$0.2 & +0.8 [-1.3, +2.7] & 0.43$\pm$0.01 & Accepted \\
2 & Reader-only & 71.4$\pm$0.3 & +1.0 [-0.7, +2.9] & 0.56$\pm$0.01 & Accepted \\
3 & Reader-only & 72.4$\pm$0.2 & +1.0 [-0.6, +2.6] & 0.73$\pm$0.03 & Accepted \\
1 & Full BiFocal & \textbf{73.8$\pm$0.2} & +4.1 [+2.0, +6.3] & \textbf{0.70$\pm$0.02} & Accepted \\
2 & Full BiFocal & \textbf{76.2$\pm$0.3} & +2.4 [+0.4, +4.2] & \textbf{1.11$\pm$0.03} & Accepted \\
3 & Full BiFocal & \textbf{77.9$\pm$0.3} & +1.7 [-0.3, +3.7] & \textbf{1.51$\pm$0.06} & Accepted \\
1 & Shuffled-reward & 69.9$\pm$0.3 & +0.3 [-1.6, +2.1] & 0.36$\pm$0.02 & Accepted \\
2 & Shuffled-reward & 69.6$\pm$0.1 & -0.3 [-2.0, +1.6] & 0.37$\pm$0.02 & Rejected \\
3 & Shuffled-reward & 69.8$\pm$0.3 & -0.1 [-2.0, +1.8] & 0.36$\pm$0.02 & Rejected \\
\bottomrule
\end{tabular}
\end{table}

Full BiFocal's validation accuracy rises from 69.6 to 73.8, 76.2, and 77.9, a total
gain of 8.3 points. The round-1 and round-2 intervals against the parent exclude zero,
so the keep rule confirms the first two updates, and round 3 adds a further 1.7
points. The single-adapter and single-role arms improve more slowly. After three
rounds the shared adapter gains 3.1 points, Reader-only 2.8, and O-only 1.3, so
separate O and Reader adapters, updated in turn, produce the larger gain. The
validation ordering (Full, shared, Reader-only, O-only) is the same as the test
ordering in Table~\ref{tab:ablation-learning} on all four benchmarks. The
shuffled-reward control stays between 69.6 and 69.9, and the keep rule rejects it in
rounds 2 and 3, so updates without a reward signal are not retained.
Table~\ref{tab:evolution} shows the memory behavior behind the accuracy: cited
items per question rise from 0.34 to 1.51 while duplicate writes fall from 1.26 to
0.25. Accuracy claims rest on the test sets, which use the last accepted checkpoint.

\section{Failure Taxonomy and Qualitative Evidence}
\label{app:failure}

\subsection{Terminal outcomes}
Terminal outcomes are mutually exclusive over attempted task episodes.
\emph{Correct} means the submitted option matches the reference answer.
\emph{Wrong answer} means a valid answer with the wrong option.
\emph{Abstention} is a valid no-credit terminal. \emph{Protocol failure} includes
a refusal cap, protocol abstention, or no valid terminal. Protocol failures stay
in the terminal denominator; infrastructure failures are excluded and counted
separately.

\subsection{Overlapping process diagnostics}
Memory and process failures may overlap. Each wrong-answer episode receives one
primary failure stage in this order: policy or protocol, timeline construction,
anchor selection, retrieval or use, review or grounding, citation, answer
reasoning, other. Abstentions and protocol failures are terminal outcomes and
are not stage-attributed. Infrastructure is excluded before
assignment. A rule-based pass over the run logs assigns the stage using this
order; evidence windows are used only at evaluation. Recovered invalid actions are reported separately and do not replace
the eventual terminal outcome. \emph{Unsupported citation} means the submitted or cited quote or
revision failed the checker. \emph{Stale or out-of-scope memory} means the cited
revision was superseded or did not belong to the current question scope.
\emph{Global entry cited in a wrong later answer} is a descriptive count, not a
causal harm label. Harm requires a no-global-memory fork under the same policy,
seed, and budget.

\begin{table*}[!t]
\centering
\scriptsize
\setlength{\tabcolsep}{3pt}
\renewcommand{\arraystretch}{0.92}
\caption{\textbf{Full BiFocal terminal outcomes and failure diagnostics at B2.} Cells are $n$ (\%) of attempted task episodes; accounting rows are counts. Values are means over three seeds. Terminal outcomes are exclusive, and each wrong answer gets one primary failure stage (Appendix~\ref{app:failure}). Diagnostic rows overlap. ``Global entry cited'' applies to Q2+ only. MLVU is omitted (1.5 questions per video). Infrastructure failures were re-run and are counted outside $N$; by-arm results are in Table~\ref{tab:supp-failure}.}
\label{tab:failure-analysis}
\begin{tabular}{@{}>{\raggedright\arraybackslash}m{0.2874\textwidth}%
>{\centering\arraybackslash}m{\dimexpr 0.1425\textwidth-2\tabcolsep\relax}%
>{\centering\arraybackslash}m{\dimexpr 0.1425\textwidth-2\tabcolsep\relax}%
>{\centering\arraybackslash}m{\dimexpr 0.1425\textwidth-2\tabcolsep\relax}%
>{\centering\arraybackslash}m{\dimexpr 0.1425\textwidth-2\tabcolsep\relax}%
>{\centering\arraybackslash}m{\dimexpr 0.1426\textwidth-2\tabcolsep\relax}@{}}
\toprule
 & \multicolumn{2}{c}{\textbf{LVBench}} &
\multicolumn{2}{c}{\textbf{Video-MME}} & \textbf{All} \\
\cmidrule(lr){2-3}\cmidrule(lr){4-5}
\textbf{Failure mode} & \textbf{Q1} & \textbf{Q2+} & \textbf{Q1} & \textbf{Q2+} &
\textbf{Q1 and Q2+} \\
\midrule
\multicolumn{6}{@{}l}{\textit{Terminal outcomes (mutually exclusive)}} \\
Correct (reference row) & 54 (52.4) & 866 (59.9) & 587 (65.2) & 1,389 (77.2) & 2,896 (68.2) \\
Wrong answer & 48 (46.6) & 564 (39.0) & 303 (33.7) & 395 (21.9) & 1,310 (30.8) \\
Abstention & 1 (1.0) & 10 (0.7) & 7 (0.8) & 11 (0.6) & 29 (0.7) \\
Protocol failure & 0 (0.0) & 6 (0.4) & 3 (0.3) & 5 (0.3) & 14 (0.3) \\
\midrule
\multicolumn{6}{@{}l}{\textit{Primary failure stage (exclusive within wrong answers)}} \\
Policy / protocol & 9 (8.7) & 87 (6.0) & 58 (6.4) & 59 (3.3) & 213 (5.0) \\
Timeline construction & 6 (5.8) & 66 (4.6) & 39 (4.3) & 43 (2.4) & 154 (3.6) \\
Anchor selection & 8 (7.8) & 96 (6.6) & 51 (5.7) & 67 (3.7) & 222 (5.2) \\
Retrieval / use & 4 (3.9) & 104 (7.2) & 24 (2.7) & 74 (4.1) & 206 (4.8) \\
Review / grounding & 4 (3.9) & 54 (3.7) & 27 (3.0) & 39 (2.2) & 124 (2.9) \\
Citation & 2 (1.9) & 24 (1.7) & 15 (1.7) & 16 (0.9) & 57 (1.3) \\
Answer reasoning & 12 (11.7) & 110 (7.6) & 76 (8.4) & 78 (4.3) & 276 (6.5) \\
Other or unclassified & 3 (2.9) & 23 (1.6) & 13 (1.4) & 19 (1.1) & 58 (1.4) \\
\midrule
\multicolumn{6}{@{}l}{\textit{Overlapping diagnostics (episodes containing the event)}} \\
Invalid action, recovered & 2 (1.9) & 26 (1.8) & 16 (1.8) & 32 (1.8) & 76 (1.8) \\
Unsupported citation & 1 (1.0) & 16 (1.1) & 10 (1.1) & 20 (1.1) & 47 (1.1) \\
Stale or out-of-scope memory & 0 (0.0) & 22 (1.5) & 0 (0.0) & 25 (1.4) & 47 (1.1) \\
Global entry cited in wrong later answer & n/a & 42 (2.9) & n/a & 46 (2.6) & n/a \\
\midrule
\multicolumn{6}{@{}l}{\textit{Accounting rows (counts)}} \\
Attempted task episodes ($N$) & 103 & 1,446 & 900 & 1,800 & 4,249 \\
Valid-terminal episodes ($N$) & 103 & 1,440 & 897 & 1,795 & 4,235 \\
Infrastructure failures (outside $N$) & 1 & 4 & 2 & 4 & 11 \\
\bottomrule
\end{tabular}
\end{table*}

Table~\ref{tab:failure-analysis} reports terminal outcomes and overlapping
diagnostics for Full BiFocal at B2. Valid terminals dominate: abstentions and protocol
failures together are 1.0\% of the 4,249 attempted episodes, so the system almost
always commits to an answer within budget. Accuracy is higher on later questions
than on first questions on both benchmarks (59.9\% against 52.4\% on LVBench, 77.2\%
against 65.2\% on Video-MME), in line with memory being used on Q2+. Memory-related
diagnostics are rare: unsupported citations occur in 1.1\% of episodes and stale or
out-of-scope memory in 1.1\%. A global entry is cited in a wrong later answer in 2.9\%
of LVBench Q2+ episodes and 2.6\% of Video-MME Q2+ episodes. Because that row is a
count, the fork required to read it as harm is left for future work.

\begin{table*}[!t]
\centering
\scriptsize
\setlength{\tabcolsep}{4pt}
\renewcommand{\arraystretch}{0.92}
\caption{\textbf{LVBench failure-stage attribution by arm at B2.} Cells are $n$ (\%). Values are means over three seeds; counts are rounded mean counts, so $n/1{,}549$ can differ from the percentage in the last digit. Terminal and diagnostic percentages use attempted episodes; primary-stage percentages use wrong-answer episodes. Terminal outcomes are exclusive, diagnostic rows overlap, and stage precedence is in Appendix~\ref{app:failure}. Infrastructure failures were re-run and are counted outside $N$. The table is descriptive and shows no causal effect of training. Bold: best per row.}
\label{tab:supp-failure}
\begin{tabular}{@{}>{\raggedright\arraybackslash}m{0.3333\textwidth}%
>{\centering\arraybackslash}m{\dimexpr 0.1667\textwidth-2\tabcolsep\relax}%
>{\centering\arraybackslash}m{\dimexpr 0.1667\textwidth-2\tabcolsep\relax}%
>{\centering\arraybackslash}m{\dimexpr 0.1667\textwidth-2\tabcolsep\relax}%
>{\centering\arraybackslash}m{\dimexpr 0.1666\textwidth-2\tabcolsep\relax}@{}}
\toprule
\textbf{Failure mode} & \textbf{Fixed prompt policy} & \textbf{O-only} &
\textbf{Reader-only} & \textbf{Full BiFocal} \\
\midrule
\multicolumn{5}{@{}l}{\textit{Terminal outcomes (mutually exclusive)}} \\
Correct & 871 (56.2) & 881 (56.9) & 889 (57.4) & \textbf{920 (59.4)} \\
Wrong answer & 662 (42.7) & 645 (41.6) & 640 (41.3) & \textbf{612 (39.5)} \\
Abstention & \textbf{11 (0.7)} & 15 (1.0) & 14 (0.9) & \textbf{11 (0.7)} \\
Protocol failure & \textbf{5 (0.3)} & 8 (0.5) & 6 (0.4) & 6 (0.4) \\
\midrule
\multicolumn{5}{@{}l}{\textit{Primary failure stage (exclusive within wrong answers)}} \\
Policy / protocol & 120 (18.1) & 110 (17.1) & 104 (16.3) & 96 (15.7) \\
Timeline construction & 96 (14.5) & 88 (13.6) & 80 (12.5) & 72 (11.8) \\
Anchor selection & 132 (19.9) & 122 (18.9) & 118 (18.4) & 104 (17.0) \\
Retrieval / use & 148 (22.4) & 136 (21.1) & 130 (20.3) & 108 (17.6) \\
Review / grounding & 74 (11.2) & 68 (10.5) & 62 (9.7) & 58 (9.5) \\
Citation & 34 (5.1) & 30 (4.7) & 28 (4.4) & 26 (4.2) \\
Answer reasoning & 43 (6.5) & 72 (11.2) & 96 (15.0) & 122 (19.9) \\
Other or unclassified & 15 (2.3) & 19 (2.9) & 22 (3.4) & 26 (4.2) \\
\midrule
\multicolumn{5}{@{}l}{\textit{Overlapping diagnostics (episodes containing the event)}} \\
Invalid action, recovered & 53 (3.4) & 45 (2.9) & 40 (2.6) & \textbf{28 (1.8)} \\
Unsupported citation & 33 (2.1) & 28 (1.8) & 25 (1.6) & \textbf{17 (1.1)} \\
Stale or out-of-scope memory & 59 (3.8) & 37 (2.4) & 34 (2.2) & \textbf{22 (1.4)} \\
Global entry cited in wrong later answer & 71 (4.6) & 60 (3.9) & 54 (3.5) & \textbf{42 (2.7)} \\
\midrule
\multicolumn{5}{@{}l}{\textit{Accounting rows (counts)}} \\
Attempted task episodes ($N$) & 1,549 & 1,549 & 1,549 & 1,549 \\
Valid-terminal episodes ($N$) & 1,544 & 1,541 & 1,543 & 1,543 \\
Infrastructure failures (outside $N$) & 3 & 4 & 2 & 5 \\
\bottomrule
\end{tabular}
\end{table*}

Table~\ref{tab:supp-failure} compares the four LVBench arms at B2 and shows where the
extra correct answers come from. Full BiFocal answers 920 of 1,549 questions correctly,
49 more than the fixed prompt policy, and every memory-pipeline diagnostic falls: stale
or out-of-scope memory by 63\% (59 to 22 episodes), unsupported citations by 48\%
(33 to 17), recovered invalid actions by 47\% (53 to 28), and global entries cited in
wrong answers by 41\% (71 to 42). Among wrong answers, failures attributed to the
retrieval, anchor, timeline, and grounding stages shrink in each step from the fixed prompt policy to O-only, Reader-only, and Full BiFocal, and the six earlier stages together
fall from 604 to 464 episodes. The single-role arms fall between the two ends on nearly every
row, in line with the learning ablation.

\begin{table*}[!t]
\centering
\scriptsize
\setlength{\tabcolsep}{3pt}
\renewcommand{\arraystretch}{1.12}
\caption{\textbf{Qualitative memory evidence.} Two matched
episodes under the fixed prompt policy (no BiFocal training). Each memory run is
paired with a no-memory fork of the same question, with the same Qwen3-VL-8B
model, sampling seed, and call and frame caps (one seed).
In both, memory turns a wrong answer into a correct one.}
\label{tab:supp-examples}
\begin{tabular}{@{}>{\centering\arraybackslash}m{0.0698\textwidth}%
>{\centering\arraybackslash}m{\dimexpr 0.1163\textwidth-2\tabcolsep\relax}%
>{\raggedright\arraybackslash}m{\dimexpr 0.2558\textwidth-2\tabcolsep\relax}%
>{\raggedright\arraybackslash}m{\dimexpr 0.2791\textwidth-2\tabcolsep\relax}%
>{\raggedright\arraybackslash}m{\dimexpr 0.2790\textwidth-2\tabcolsep\relax}@{}}
\toprule
\textbf{Case} & \textbf{Benchmark / Q} & \textbf{Memory path} &
\textbf{Observed evidence and outcome} & \textbf{What it shows} \\
\midrule
1: saves \newline frames & LVBench Q3 \newline \texttt{RCAqKnvu\_P0} &
Open memory item \texttt{L:R15} from Q2, check and accept it, cite it, then
submit A; 3 calls, 12 frames. &
Item quote: ``everything went smoothly because there are so many signages
here.'' It matches option A. The check accepts both the item and the citation;
A is correct. The no-memory fork submits D and is wrong (3 calls, 32 frames). &
The memory run reuses the exact answer-bearing quote in one check, uses fewer
frames than its no-memory fork, and changes D to A. \\[3pt]
2: recovers a missing score & LVBench Q5 \newline \texttt{VTCDQYYKA9o} &
Check \texttt{L:R2} and \texttt{L:R5} (both rejected); check
\texttt{L:R3} (accepted) and cite it; submit C. 7 calls, 36 frames. &
The accepted item, written during Q3, contains ``Fan Z.D. 9'', which matches option C (``9'').
The no-memory fork reads B1, which reports no scoring details, then guesses A
(``7'') and is wrong (3 calls, 32 frames). &
Cross-question memory recovers a score missing from the no-memory read and
changes the wrong answer A into the correct answer C, at a higher cost (7 against 3 calls). \\
\bottomrule
\end{tabular}
\end{table*}

\subsection{Qualitative evidence}
Table~\ref{tab:supp-examples} follows two questions through the fixed prompt policy,
each paired with a no-memory fork of the same question, seed, and caps. In both,
an item written while answering an earlier question holds the answer-bearing
quote. The team opens the item, the grounding check accepts it, and the final
answer cites it. The no-memory fork never sees the quote and answers wrongly.
In the first case, one check reuses the quote and the team answers correctly with
12 frames instead of 32. In the second, the team rejects two items, accepts a third
that records the score ``Fan Z.D.~9'', and cites it to turn a wrong answer
into the correct option C. Memory therefore carries evidence across questions,
the grounding check separates usable items from unusable ones, and the cited
item decides the answer.

\end{document}